\documentclass[structabstract]{raa_twocolumn}
\usepackage{graphicx,times}
\usepackage{threeparttable}
\usepackage{natbib,amssymb}
\usepackage{float}
\usepackage{color}
\usepackage{bm}
\usepackage{subfigure}
\usepackage[figuresright]{rotating}
\usepackage{amsmath}
\usepackage{multirow}
\usepackage{url}
\usepackage[backref]{hyperref}
\usepackage{ulem}

\graphicspath{{./}{figures/}}
\begin{document}
   \title{On the identification of N-rich metal-poor field stars with future China space station telescope}
   \volnopage{ {\bf 2022} Vol.\ {\bf XX} No. {\bf XXX}, 000--000}
   \setcounter{page}{1}
   \author{Jiajun Zhang \inst{1,2}, Baitian Tang \inst{1,2}, Jiang Chang \inst{3}, Xiangxiang Xue \inst{4}, Jos\'e G. Fern\'andez-Trincado \inst{5}, Chengyuan Li \inst{1,2}, Long Wang \inst{1,2}, Hao Tian \inst{6}, Yang Huang \inst{7,6}}
   \institute{
   School of Physics and Astronomy, Sun Yat-sen University, Zhuhai 519082, China; {\it tangbt@mail.sysu.edu.cn}\\
   \and
   CSST Science Center for the Guangdong-HongKong-Macau Great Bay Area, Sun Yat-sen University, Zhuhai 519082, China;\\
   \and
   Purple Mountain Observatory, Chinese Academy of Sciences, No.8 Yuanhua Road, Qixia District, Nanjing 210034, China;\\
   \and
   Key Lab of Optical Astronomy, National Astronomical Observatories, Chinese Academy of Sciences, 20A Datun Road, Chaoyang District, Beijing 100101, China;\\
   \and
   Instituto de Astronom\'ia, Universidad Cat\'olica del Norte, Av. Angamos 0610, Antofagasta, Chile;\\
   \and
   Key Lab of Space Astronomy and Technology, National Astronomical Observatories, Chinese Academy of Sciences, Beijing 100101, China;
   \and
   School of Astronomy and Space Science, University of Chinese Academy of Sciences, Beijing 100049, China
}  
   \date{Received; accepted}

\abstract{During the long term evolution of globular clusters (GCs), a part of member stars are lost to the field. The recently found nitrogen-rich (N-rich) metal-poor field stars are promising candidates of these GC escapees, since N enhancement is the fingerprint of chemically enhanced populations in GCs. In this work, we discuss the possibility of identifying N-rich metal-poor field stars with the upcoming China space station telescope (CSST). We focus on the main survey camera with \emph{NUV, u, g, r, i, z, y} filters and slitless spectrograph with a resolution about 200. The combination of \emph{UV} sensitive equipment and prominent N-related molecular lines in the \emph{UV} band bodes well for the identification: the color-color diagram of (\emph{u-g}) versus (\emph{g-r}) is capable of separating N-rich field stars and normal halo stars, if metallicity can be estimated without using the information of \emph{u}-band photometry. Besides, the synthetic spectra show that a signal-to-noise ratio of 10 is sufficient to identify N-rich field stars. In the near future, a large sample of N-rich field stars found by CSST, combined with state-of-the-art N-body simulations will be crucial to decipher the GC-Galaxy co-evolution.
 }
\authorrunning{Zhang et al.}
\titlerunning{N-rich field stars with CSST}
\maketitle

\section{Introduction} \label{sec:intro}
According to most recent survey of globular clusters (GCs), almost all exhibit the phenomenon of multiple stellar populations (MPs, \citealt{Gratton2012}, \citealt{Schiavon2017a}, \citealt{Nataf2019}, \citealt{Masseron2019}, \citealt{Meszaros2020}). A feature of MPs is chemical inhomogeneity of member stars, which is shown via an anti-correlation between nitrogen (N) and carbon (C), or an anti-correlation between sodium (Na) and oxygen (O), or an anti-correlation between aluminum (Al) and magnesium (Mg). The community usually considers this as a consequence of different level of enhancement of the primordial population, and ``first generation''(FG) and ``second generation''(SG) are referred to primordial and enriched populations, respectively. Typical SG stars exhibit a different chemical abundance pattern than normal field stars in a given metallicity, with enhanced N, Na, (sometimes helium (He), Al, and silicon (Si)), but depleted C, O, (sometimes Mg). 

The unique chemical fingerprint of SG stars is particularly suitable for tracing the evolution of individual stars formed in GCs. Due to the dynamical evolution of stars in GCs (including two-body relaxation and ejection) and their tidal interaction with Milky Way (MW), GC stars are lost to the field \citep[e.g.,][]{Weatherford2023}. Identifying these GC escapees in the field is crucial to evaluate different stellar escape mechanisms and estimate their contribution to the MW halo. These GC escapees may not maintain their dynamical properties for long, but chemistry can be preserved along their life time. Though FG stars are almost chemically identical to halo field stars, SG stars show clear distinguishable chemical features. Among the chemical peculiarity of various light elements that we mention before, N enrichment is the most economic one to be identified, thanks to the CN molecular lines. These molecular features can be easily found in large spectroscopy surveys, even with a spectral resolution down to $\sim 1800$. Using the low-resolution optical spectra from the SEGUE survey \citep{Yanny2009} and high-resolution near-infrared spectra from the APOGEE survey \citep{Majewski2017}, a large number of N-rich field stars have been found in the MW and its satellite galaxies \citep[e.g.,][]{Martell2010,Martell2011,Martell2016,Schiavon2017b,Fernandez_Trincado2020,Fernandez_Trincado2021}. In addition, the LAMOST survey \citep{Zhao2012} with the largest number of stellar spectra available joins the quest: \cite{Tang2019,Tang2020} found $\sim100$ N-rich metal-poor field stars, which are later proved to be chemically identical to GC member stars \citep{Yu2021}.


However, a much larger sample of N-rich field stars are required to fully evaluate their escaped mechanism and the contribution of GC escapees to the MW halo. In that sense, photometric surveys or spectroscopic surveys with lower spectral resolution could be more efficient in searching for N-rich field stars. 
In the near future, the upcoming China space station telescope (CSST) will greatly increase the sample size of N-rich field stars. Its capability to obtain \emph{UV} images and spectra makes it straightforward to detect N-enrichment, due to the NH and CN molecular bands in this wavelength range. The CSST is scheduled to be launched around 2024 and will share the same orbit with the China space station. Its main survey module has a large field of view (1.1 square degree) and a high spatial resolution ($\sim$0.15”). The limiting magnitudes in the \emph{u} and \emph{g} bands are expected to be 25.4 mag and 26.3 mag (AB magnitude) for point sources \citep{Zhan2021}. The main survey module will obtain multi-band imaging data (\emph{NUV} and \emph{ugrizy}) and slitless spectra (\emph{R} $\sim 200$, \citealt{Zhan2021}) in 17,500 square degrees of the sky in its planned ten-year observation. 
There are also four affiliated instruments on board CSST: Multi-channel Imager (MCI),
Integral Field Spectrograph (IFS), Cool Planet Imaging Coronagraph (CPIC), and
Terahertz Receiver (THz). \citet{Lichengyuan2022} demonstrated the possibility to disentangle stellar populations with different abundances of He, C, N, O and Mg with MCI/CSST photometry. In this work, we will estimate the capability of identifying N-rich field stars with photometry and slitless spectra from main survey camera.

This paper is organized as follows. Section \ref{sec:method} introduced the methodology of generating a mock photometric catalogue for MW halo stars, where N-rich field stars are considered. Next, we estimate the feasibility of separating N-rich field stars from normal halo stars with the color-color diagram (CSST main survey camera filters) and slitless spectra (Section \ref{sec:results}). Section \ref{sec:discussion} and Section \ref{sec:conclusion} is devoted to discussions and conclusions, respectively.

\section{Method}\label{sec:method}
\begin{figure*}
\includegraphics[width=1.0\textwidth]{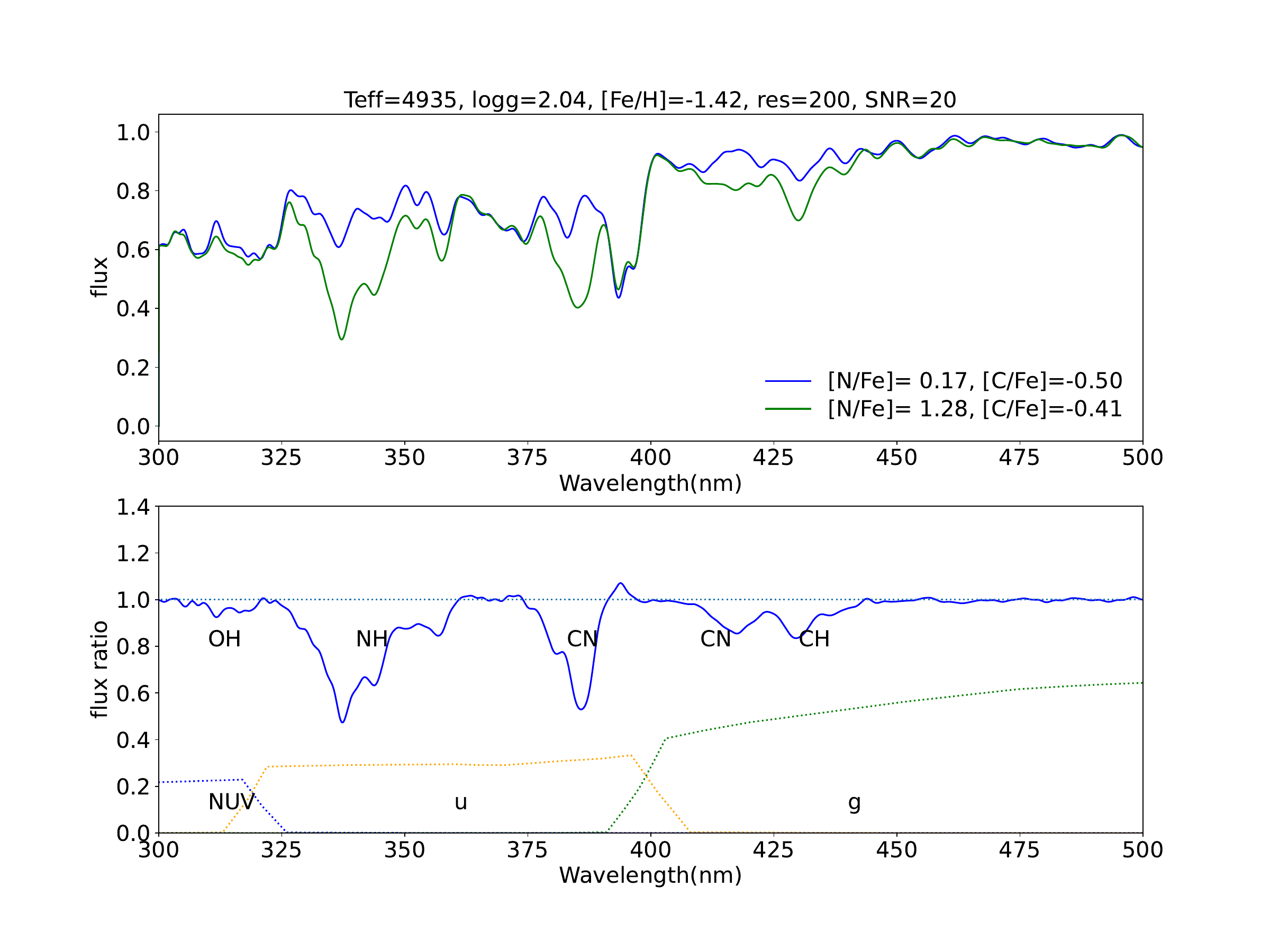}
\caption{The spectral comparison between two stars with different N enrichment. The upper panel shows two spectra of a N-rich giant star (green line) and a normal giant star (blue line). They have same atmospheric parameters ($T_{\rm eff}$, $\log g$, and [Fe/H]) but different C and N abundances. The atmospheric parameters, spectral resolution, and signal-to-noise ratio (SNR) are displayed in the figure title. The C and N abundances are listed in the bottom-right. The lower panel displays the flux ratio between the two spectra. The locations of OH3200, NH3400, CN3839, CN4142, and CH4300 are labeled. The horizontal dashed line represents a flux ratio of 1.0. The transmission curves of three CSST filters (\emph{NUV}, \emph{u}, and \emph{g}) are represented by dashed lines with different colors in the bottom. }
\label{fig:example_spectra}
\end{figure*}

The logic behind identifying N-rich stars\footnote{Since ``field'' only indicates a star is not located inside any star cluster, dropping it would not affect any other on-going projects.} with \emph{UV}-related photometry and spectra is their strong NH and CN features in this wavelength range (Figure \ref{fig:example_spectra}). The \emph{u}-band filter in the CSST main survey camera covers NH3400 and CN3839 molecular lines, while \emph{g}-band filter covers CN4142 and CH4300 molecular lines. The different profiles of molecular lines caused by peculiar C and N abundances may render a noticeable change in \emph{u}-band or \emph{g}-band magnitudes.

In order to estimate the possibility of identifying N-rich stars under a situation as realistic as possible, this work uses the smooth halo catalog\footnote{Background halo stars} from the CSST Mock Catalog of Stellar Halo (Chang et al., in prep.). The {\sc Galaxia} software \citep{Sharma2011}, which is based on the N-body simulation from \cite{Bullock2017}, is used to generate a stellar halo of the MW. The stellar halo catalogue includes six-dimensional kinematic parameters (positions and velocities), as well as information such as mass, age, metallicity, effective temperature, and luminosity for each star. Considering the footprint of CSST, the stellar halo catalog only includes stars with a galactic latitude of 25 degrees or higher. {\sc GalevNB} \citep{Pang2016} is used to generate the CSST photometric magnitudes based on these stellar parameters by convolving theoretical spectra with the CSST main camera filters\footnote{http://svo2.cab.inta-csic.es/svo/theory/fps3/}. So this catalogue also includes the photometric magnitudes in the CSST main camera filters.

In the previous catalogue, we assume solar abundance pattern with [C/Fe] = 0 and [N/Fe] = 0 for all stars. In reality, however, non-solar chemical abundances (e.g., C, N) can change the depth of feature lines in the spectra, which affects the corresponding photometric magnitudes. To obtain the magnitude differences between non-solar and solar chemical patterns, we generate the corresponding synthetic spectra ($R \sim 200$) and convolve them with the CSST main camera filters. Publicly available tool, {\sc iSpec} \citep{Blanco_Cuaresma2014, Blanco_Cuaresma2019} is employed to generate synthetic spectra, where we use the radiative transfer code  and line lists from SPECTRUM \citep{Gray1999}, MARCS model atmospheres \citep{Gustafsson2008} and \citet{Asplund2009} solar abundances. Finally, we add these magnitude differences to photometric magnitudes from the aforementioned catalogue, and we obtain a new catalogue that considers a non-solar chemical pattern.  

\section{Results} \label{sec:results}
\subsection{Photometric Identification of N-rich Giant Stars}

\begin{figure}
\center
\includegraphics[scale=0.3]{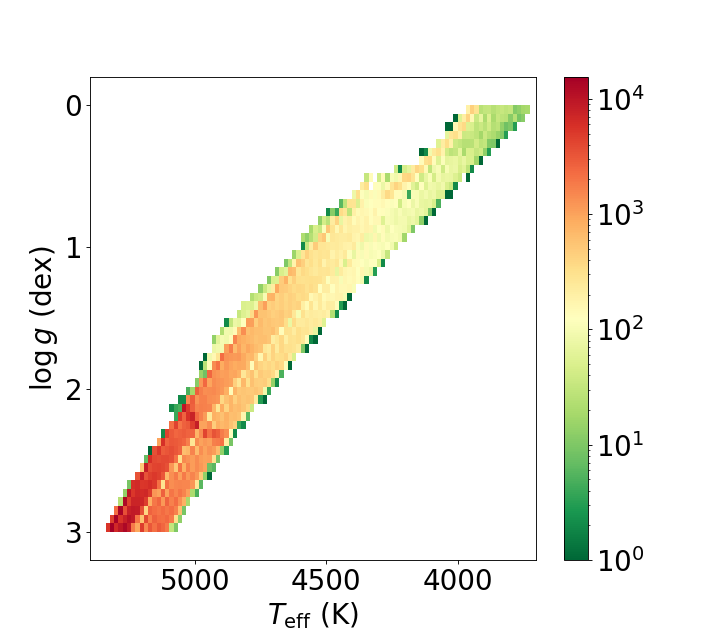}
\caption{The Kiel diagram  of the selected giant stars, with color representing the number density.  \label{fig:Kiel_giant}}
\end{figure}

Since C and N abundances for giant stars may substantially change after they go through first dredge-up and extra-mixing, we separate the giants and dwarfs in the following photometric identification of N-rich stars.
Giant stars with $\log g<$ 3.0 were selected, while blue horizontal branch stars (BHBs) on the Kiel diagram were discarded, since high temperature would destroy C and N related molecules. 
Figure \ref{fig:Kiel_giant} shows the Kiel diagram and the color indicates the number density of sample stars. Following the selection criteria and statistical results presented in \cite{Tang2020}, (1) we limited the metallicity range of our sample stars to -1.8 $<$ [Fe/H] $<$ -1.0, (2) we assumed that 1\% of the sample stars are N-rich stars and 99\% are normal stars. 

\begin{figure*}[t]
\includegraphics[width=1.0\textwidth]{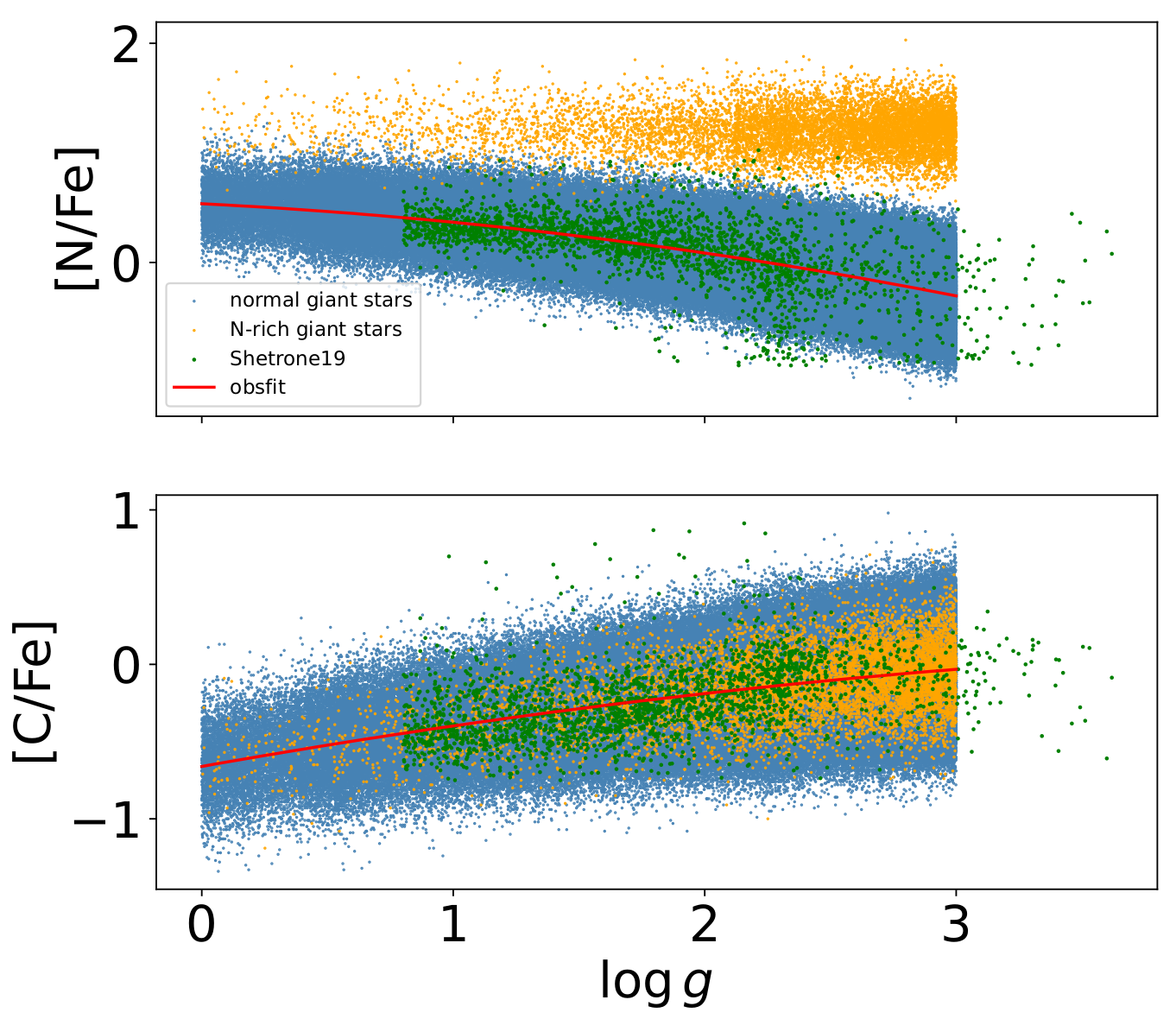}
\caption{The abundances of C and N for N-rich (orange dots) and normal giant stars (blue dots) as a function of $\log g$. Green dots represent red giant stars from \cite{Shetrone2019}, and red lines are relationships between N(C) abundances and surface gravity for the Shetrone sample. See the text for more details.}
\label{fig:fieldCN}
\end{figure*}

In order to simulate stars with real C and N abundances, we set the chemical patterns of N-rich giants and normal giants as follows: (1) normal giants. We used red giant stars from \cite{Shetrone2019} as reference and fitted the correlations between C(N) abundances and surface gravity (Figure \ref{fig:fieldCN}), since C and N abundances would change as a star climbs up red giant branch (RGB). For a star with given surface gravity, we first estimated the expected value of C and N from the fitted correlations, and then added a random error which follows a Gaussian distribution with $\mu=0$ dex, $\sigma = 0.2$ dex. 
(2) N-rich giants. The C abundance of N-rich (field) giants are yet debated, here we followed the distribution of normal giants. While for their N abundance, we followed the observational results of Tang et al. (in prep.), where we set the expected value to be 1.2 dex, and its error also follows a Gaussian distribution with $\mu=0$ dex, $\sigma = 0.2$ dex. We then generated new photometric magnitudes according to Section \ref{sec:method}.

\begin{figure}
\center
\includegraphics[scale=0.16]{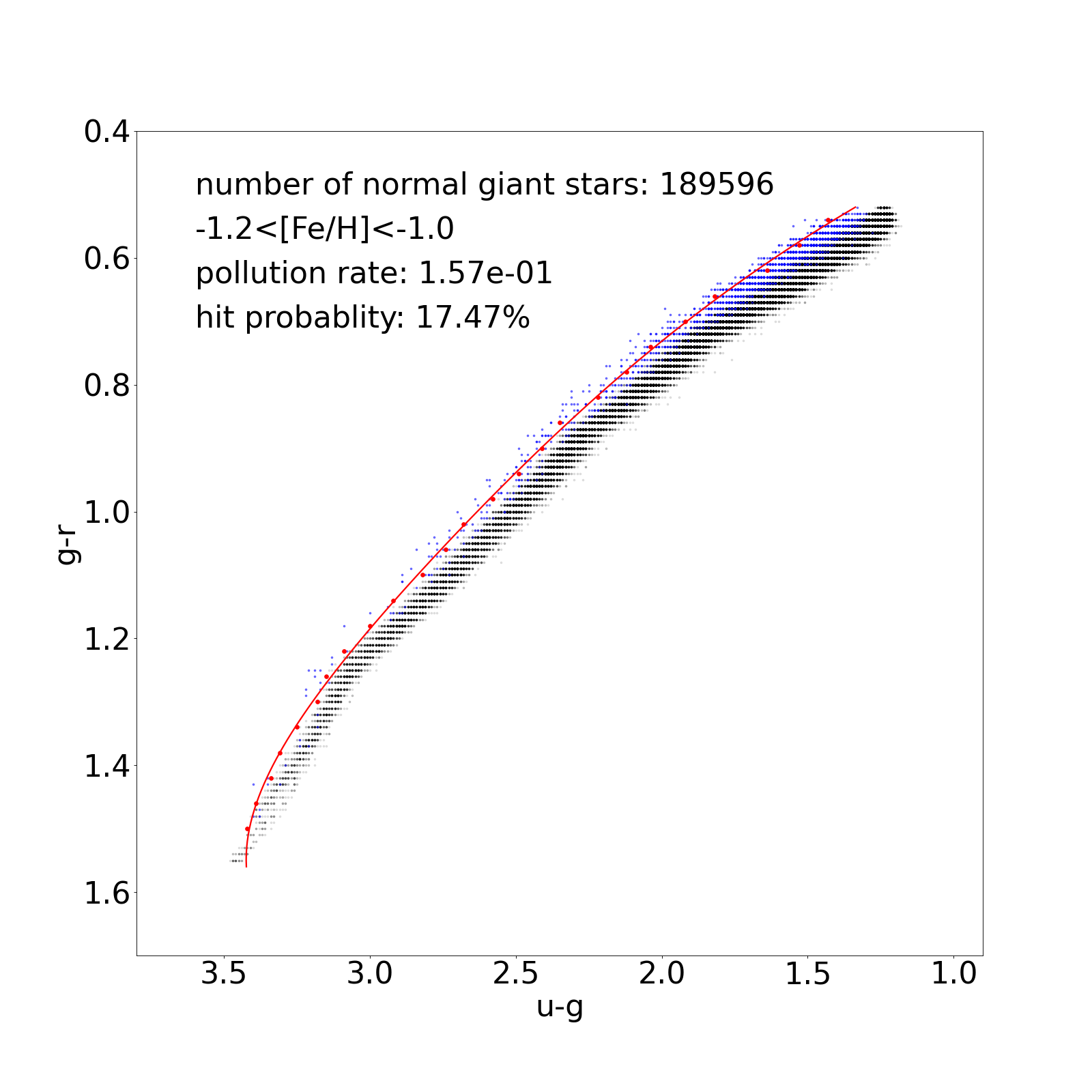}
\caption{\emph{u-g} versus \emph{g-r} for N-rich giant stars (blue) and normal giant stars (black) at -1.2 $<\rm [Fe/H]<$ -1.0. We rebinned the normal giant stars on \emph{g-r} with a bin width of 0.04 mag. We calculated the 100\% quantile in \emph{u-g} for each bin. A fifth-order polynomial fit was fitted to these quantile points (red points) and the red curve is the corresponding curve of the polynomial fit. We labeled the number of normal stars, the metallicity range, the pollution rate, and hit probability in the upper-left.\label{fig:giant_ugvsgr_feh_m12m10}}
\end{figure}

After exploring the color-magnitude diagram and color-color diagram of different combinations of CSST filters, we found that \emph{u-g} versus \emph{g-r} is a potential diagnostic color-color diagram to separate stars with different N enrichment, since we have shown that \emph{UV} and blue filters are particularly sensitive to N enrichment in Figure \ref{fig:example_spectra}. In order to identify more N-rich stars, we divided the metallicity range ([-1.8, -1.0]) into four different bins with equal bin size.
Figure \ref{fig:giant_ugvsgr_feh_m12m10} show the results in the metallicity bin of [-1.2, -1.0]\footnote{The color-color diagrams at other metallicity ranges can be found in Figure \ref{fig:nhgiant_ugvsgr_otherfeh} in Appendix.}. We rebinned the normal giant stars on \emph{g-r} with a bin width of 0.04 mag and calculated their 100\% quantile in \emph{u-g}\footnote{The maximum \emph{u-g} value for normal giant star sample.} for each bin. After obtaining the 100\% quantile points for all the \emph{g-r} bins (red points in Figure \ref{fig:giant_ugvsgr_feh_m12m10}), we fitted a fifth-order polynomial to these quantile points (red curve in Figure \ref{fig:giant_ugvsgr_feh_m12m10}). The coefficients of the polynomial is listed in Table \ref{tab:table_giant}. Stars on the left-hand side of the red curve should be pure N-rich giant stars sample. However, the real situation is slightly more complicated: the pollution rate (Table \ref{tab:table_giant}) is not zero, because (1) we binned \emph{g-r} in 0.04 step size; (2) the red curve (polynomial fit) may not follow exactly the red points. We also calculated the hit probability  (Table \ref{tab:table_giant}), which is defined as the ratio of identified N-rich stars (on the left of the red curve) to all N-rich stars. 
According to Table \ref{tab:table_giant}, the hit probabilities are large for all metallicity bins including the full metallicity bin ([-1.8, -1.0]), suggesting the color-color diagram of \emph{u-g} versus \emph{g-r} is a potential way to separate N-rich stars from normal stars.
By comparing the hit probabilities at different metallicity ranges, we find that the bins with higher metallicity ([-1.2, -1.0] and [-1.4, -1.2]) show larger values, because of stronger molecular features at higher metallicity.
The hit probabilities at divided metallicity ranges are larger than that at the full metallicity range, suggesting that subdividing the metallicity range can help us select more N-rich giant stars.

\begin{table*}
\centering
\caption{The Pollution Rate, Hit Probability, and Coefficients of Division Lines for Giant Stars without Photometric Error.}
\label{tab:table_giant}
\begin{threeparttable}
\resizebox{\textwidth}{!}{
\renewcommand\arraystretch{2}
\begin{tabular}[b]{lllllllll}
\hline
\hline
[Fe/H] bin & pollution rate & hit probability & $a_{0}$ & $a_{1}$ & $a_{2}$ & $a_{3}$ & $a_{4}$ & $a_{5}$\\
\hline
-1.2$\rm <[Fe/H]<$-1.0 & 1.57e-01 & 17.47\% & -1.790 & 8.931 & -6.535 & 1.112 & 1.653 & -0.739\\
-1.4$\rm <[Fe/H]<$-1.2 & 1.36e-02 & 24.76\% & 2.732 & -18.027 & 55.059 & -67.090 & 38.330 & -8.424\\
-1.6$\rm <[Fe/H]<$-1.4 & 1.51e-02 & 10.55\% & 3.937 & -25.291 & 72.353 & -87.648 & 50.343 & -11.173\\
-1.8$\rm <[Fe/H]<$-1.6 & 9.68e-03 & 15.14\% & 4.796 & -30.073 & 81.638 & -95.357 & 52.768 & -11.272\\
-1.8$\rm <[Fe/H]<$-1.0 & 3.95e-02 & 7.72\% & -2.549 & 13.164 & -16.203 & 12.090 & -4.448 & 0.580\\
\hline
\end{tabular}}
\end{threeparttable}
\raggedright{\hspace{2cm}Note: $y = a_{0} + a_{1}x + a_{2}x^{2} + a_{3}x^{3} + a_{4}x^{4} + a_{5}x^{5}$, $x = $ \emph{g - r}, $y = $ \emph{u - g}.\\}
\end{table*}

\subsection{Photometric Identification of N-rich Dwarf Stars}

\begin{figure}
\center
\includegraphics[scale=0.3]{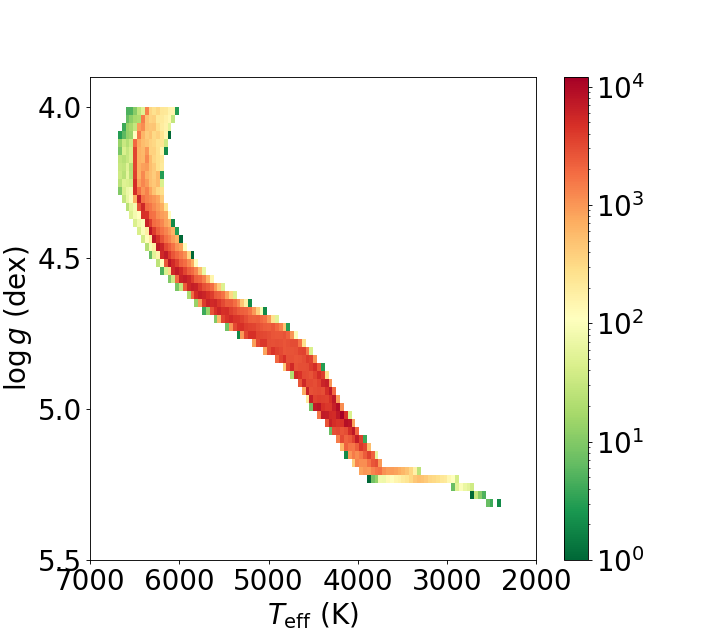}
\caption{The Kiel diagram of selected dwarf stars, with color representing the number density.  \label{fig:Kiel_dwarf}}
\end{figure}

Due to the vast number of dwarf stars in our catalogue, we randomly drew 1\% out of the dwarf star sample ($\log g>$ 4.0, -1.8 $<$ [Fe/H] $<$ -1.0) to save computational time. Given that the capability of identifying N-rich stars would not be affected by the sample size after it reaches a statistically significant number, we consider our simplified dwarf star sample ($\sim 10^{6}$) sufficient for the following discussion. The Kiel diagram (Figure \ref{fig:Kiel_dwarf}) shows that our sample covers the expected $T_{\rm eff}$ and $\log g$ parameter space.
For N-rich dwarf stars (1\% of our simplified dwarf star sample), their [N/Fe] are set as expected value of 0.8 dex plus a random error which follows a Gaussian distribution with $\mu=0$ dex, $\sigma=0.2$ dex, and their [C/Fe] are set as expected value of -0.26 dex plus a random error which follows a Gaussian distribution with $\mu=0$ dex, $\sigma=0.2$ dex. Then, we calculated their new photometric magnitudes according to section \ref{sec:method}. Since normal dwarf stars (the other 99\%) are usually not chemically enhanced, solar chemical pattern is assumed, and no changes were made to their photometric magnitudes. 

\begin{figure}
\center
\includegraphics[scale=0.16]{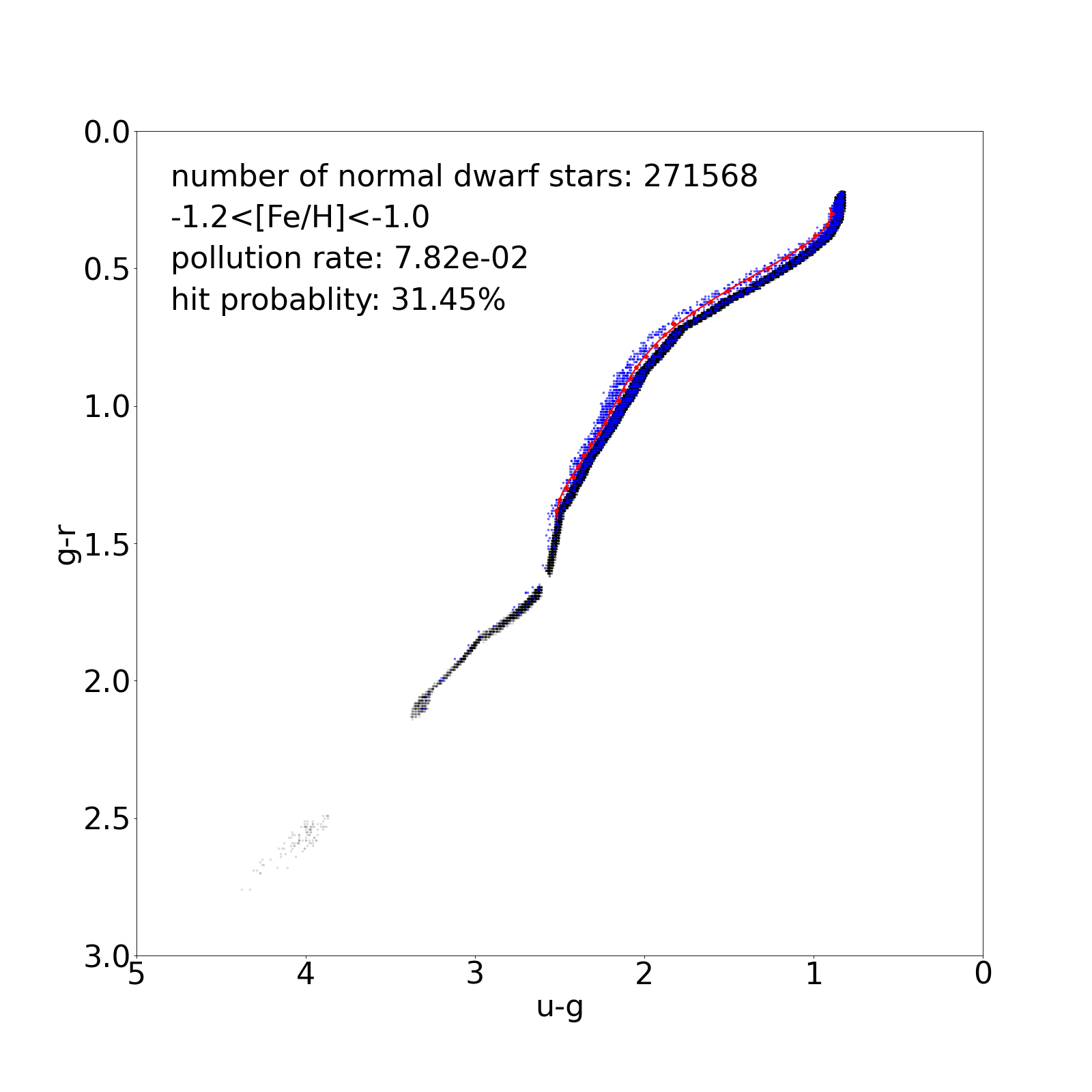}
\caption{\emph{u-g} versus \emph{g-r} for N-rich dwarf stars (blue) and normal dwarf stars (black) at -1.2 $<\rm [Fe/H]<$ -1.0. Symbol meanings are the same as Figure \ref{fig:giant_ugvsgr_feh_m12m10}. The red curve ranges from \emph{g-r} = 0.28 mag to 1.40 mag.\label{fig:ugvsgr_feh_m12m10}}
\end{figure}

Figure \ref{fig:ugvsgr_feh_m12m10} shows \emph{u-g} versus \emph{g-r} color-color diagram of N-rich dwarf stars and normal dwarf stars at the metallicity bin of [-1.2, -1.0]\footnote{The color-color diagrams at other metallicity ranges could be found in Figure \ref{fig:ugvsgr_otherfeh} in Appendix.}. Following similar procedure as the giant stars, we obtained the border between two populations (red curve) in Figure \ref{fig:ugvsgr_feh_m12m10}. However, this border was obtained only between the \emph{g-r} = 0.28-1.40 mag (same for all the metallicity bins), because (1) the two populations have significant overlap at \emph{g-r} $<$ 0.28 mag, and it is difficult to separate them. (2) the fitting is significantly worse if it is extended to \emph{g-r} $>$ 1.40 mag, which corresponds to the less constrained ``knee'' feature in the lower main sequence. We showed the pollution rate, hit probability, and coefficients of division lines at different metallicity ranges in Table \ref{tab:table_dwarf}. According to Table \ref{tab:table_dwarf}, for all metallicity bins including the full metallicity bin ([-1.8, -1.0]), the hit probabilities are large and pollution rates are very small, suggesting that the color-color diagram of \emph{u-g} versus \emph{g-r} can separate N-rich stars from normal stars with a high efficiency. The hit probabilities for each metallicity bin are larger than that for full metallicity bin --- dividing the metallicity range can help us select more N-rich dwarf stars.

\begin{table*}
\centering
\caption{The Pollution Rate, Hit Probability, and Coefficients of Division Lines for Dwarf Stars without Photometric Error.}
\label{tab:table_dwarf}
\begin{threeparttable}
\resizebox{\textwidth}{!}{
\renewcommand\arraystretch{2}
\begin{tabular}[b]{lllllllll}
\hline
\hline
[Fe/H] bin & pollution rate & hit probability & $a_{0}$ & $a_{1}$ & $a_{2}$ & $a_{3}$ & $a_{4}$ & $a_{5}$\\
\hline
-1.2$\rm <[Fe/H]<$-1.0 & 7.82e-02 & 31.45\% & 3.750 & -24.691 & 73.863 & -92.609 & 53.678 & -11.812\\
-1.4$\rm <[Fe/H]<$-1.2 & 9.41e-02 & 29.47\% & 3.680 & -24.571 & 73.921 & -93.185 & 54.305 & -12.012\\
-1.6$\rm <[Fe/H]<$-1.4 & 8.35e-02 & 30.20\% & 3.659 & -24.817 & 75.092 & -95.292 & 55.907 & -12.448\\
-1.8$\rm <[Fe/H]<$-1.6 & 6.84e-02 & 35.62\% & 3.813 & -26.336 & 79.847 & -102.050 & 60.359 & -13.549\\
-1.8$\rm <[Fe/H]<$-1.0 & 5.44e-02 & 13.01\% & 3.750 & -24.691 & 73.863 & -92.609 & 53.678 & -11.812\\
\hline
\end{tabular}}
\end{threeparttable}
\raggedright{\hspace{2cm}Note: $y = a_{0} + a_{1}x + a_{2}x^{2} + a_{3}x^{3} + a_{4}x^{4} + a_{5}x^{5}$, $x = $ \emph{g - r}, $y = $ \emph{u - g}.\\}
\end{table*}

\subsection{Identification of N-rich Stars with Slitless Spectrograph}

Slitless spectrograph ($ R \sim$ 200) in the main survey module of the CSST covers several strong C, N, O related molecular features, and thus should be suitable for separating N-rich stars from normal stars (see Section \ref{sec:method}). To evaluate the required SNR for observation, we synthesized spectra with $ R \sim$ 200 using {\sc iSpec}. Then Poisson noise was added to the synthetic spectra to simulate CSST observed spectra with SNR of 20/10. 

\begin{figure*}
    \centering
    \subfigure[SNR = 20]{
    \label{fig:giant_spectral_signal_SNR20}
    \includegraphics[width=0.8\textwidth]{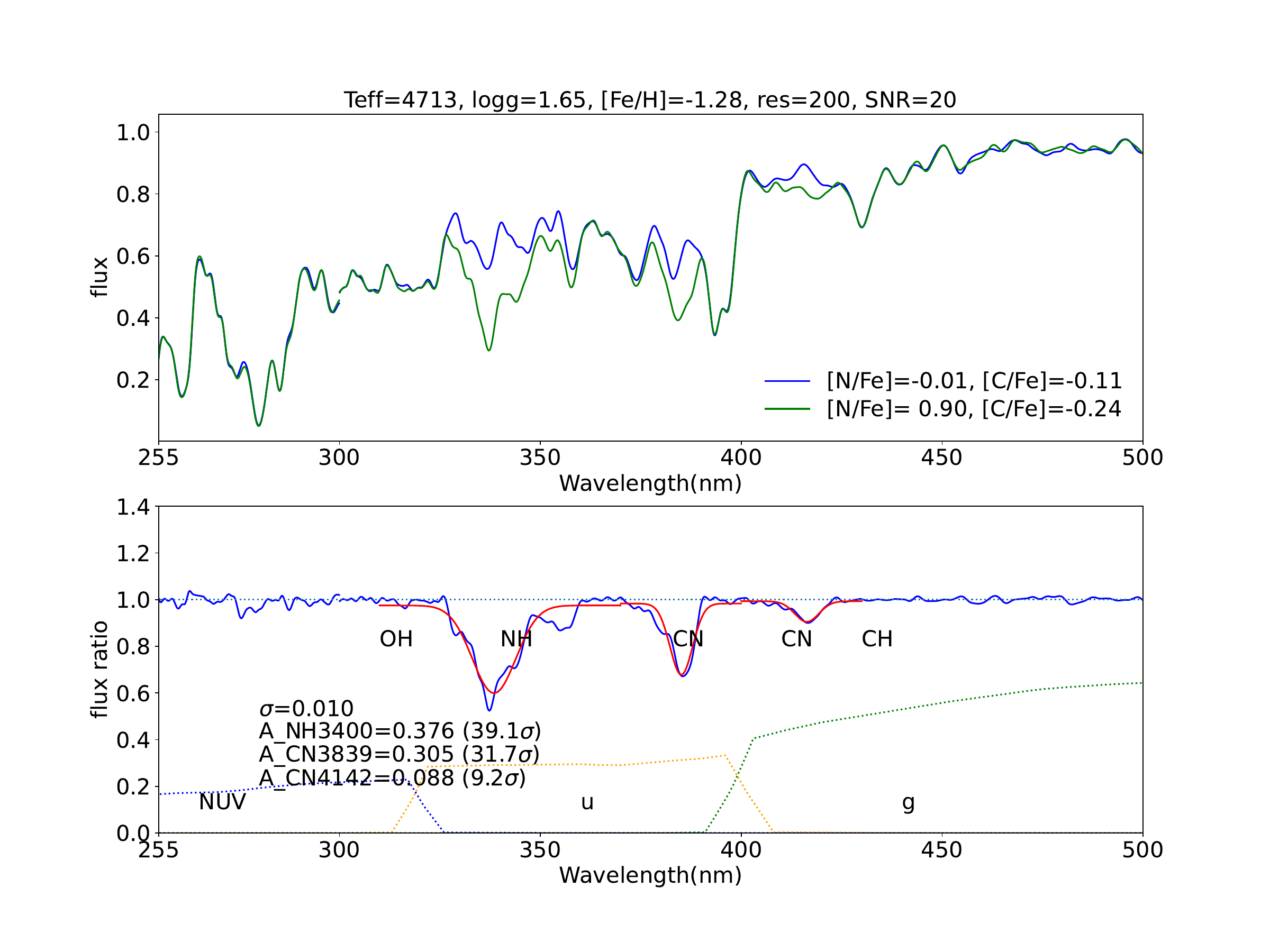}}
    \subfigure[SNR = 10]{
    \label{fig:giant_spectral_signal_SNR10}
    \includegraphics[width=0.8\textwidth]{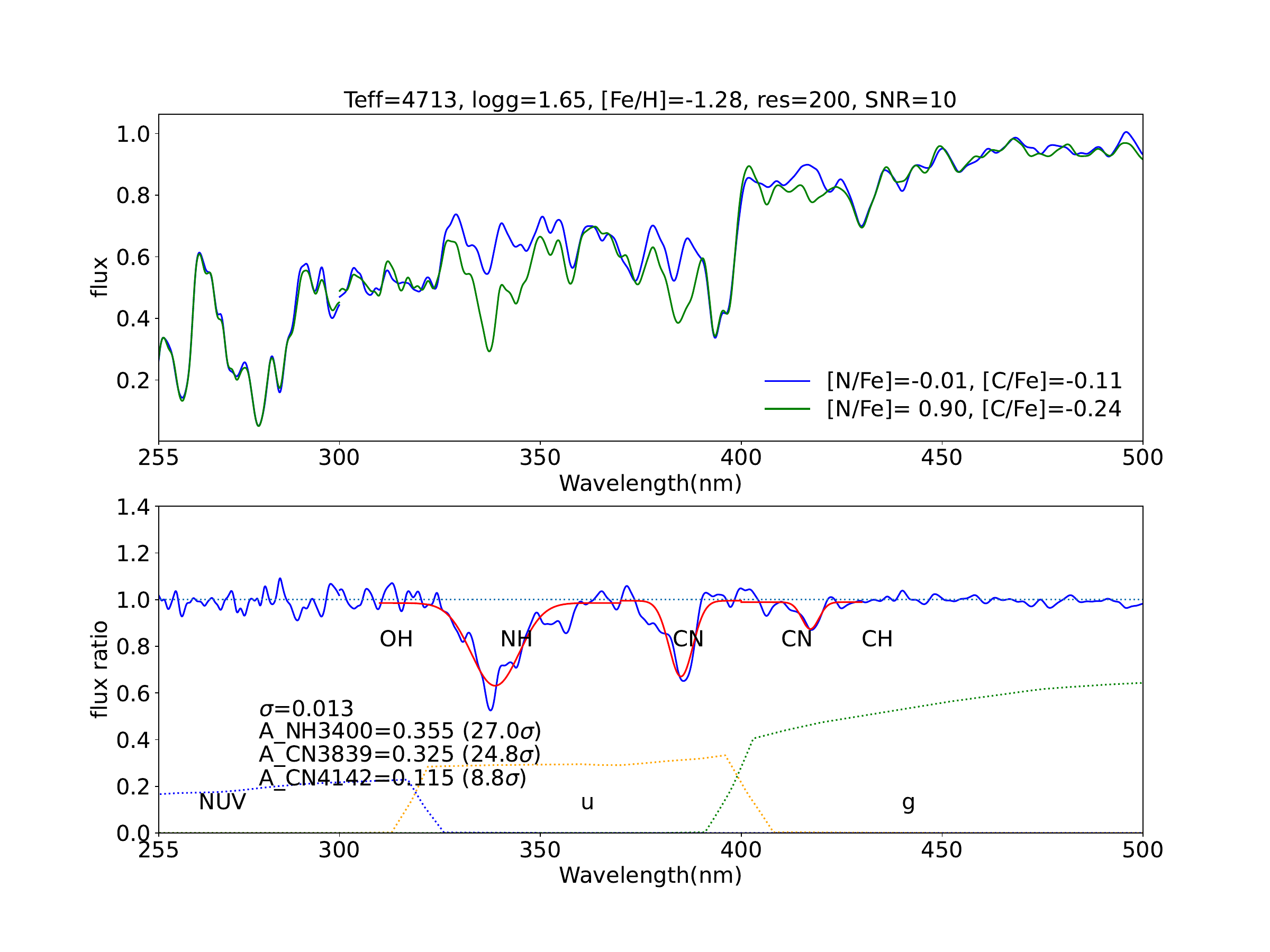}}
    \caption{Synthetic spectra with SNR of 20 (a) and 10 (b) for a representative giant star. Symbols have the same meanings as Figure \ref{fig:example_spectra}. Additionally, we calculated the significant level of strong molecular features, and listed their values in the bottom-left. See the text for more details.}
    \label{fig:giant_spectral_signal}
\end{figure*}

We picked one star with stellar parameters around the median values of the giant star sample as representative. Two spectra following the chemical patterns of N-rich stars (green lines) and normal stars (blue lines) were generated, respectively (e.g., upper panel of Figure \ref{fig:giant_spectral_signal_SNR20}. To clearly visualize the molecular features, we calculated the flux ratio between the two spectra aforementioned (e.g., lower panel of Figure \ref{fig:giant_spectral_signal_SNR20}. To further estimate their significant level, we fitted Gaussian profile (red lines) to the molecular bands in the flux ratio spectra. We used the maximum depth of the Gaussian function to represent the ``signal'' of a molecular band (labeled as ``A'' ). Then the ``noise'' ($\sigma$) is calculated as the standard deviation of flux ratios from 450 nm to 500 nm. Finally, we define the ratio of ``signal'' to ``noise'' as the significant level of each feature.
Three molecular features stand out clearly: NH3400, CN3839, and CN4142. Under an SNR of 20, the significant levels of NH3400, CN3839, and CN4142 are 39.1$\sigma$, 31.7$\sigma$, and 9.2$\sigma$, respectively, which are labeled in the bottom-left in lower panel of Figure \ref{fig:giant_spectral_signal_SNR20}. Under an SNR of 10, the significant levels of these molecular bands are 27.0$\sigma$, 24.8$\sigma$, and 8.8$\sigma$. As the SNR decreases, the significant levels of the molecular bands decrease. The significant levels are larger than 5$\sigma$ under an SNR of 10, suggesting that a spectrum with an SNR of 10 is sufficient for separating N-rich giant stars from normal giant stars.

\begin{figure*}
    \centering
    \subfigure[SNR = 20]{
    \label{fig:dwarf_spectral_signal_SNR20}
    \includegraphics[width=0.8\textwidth]{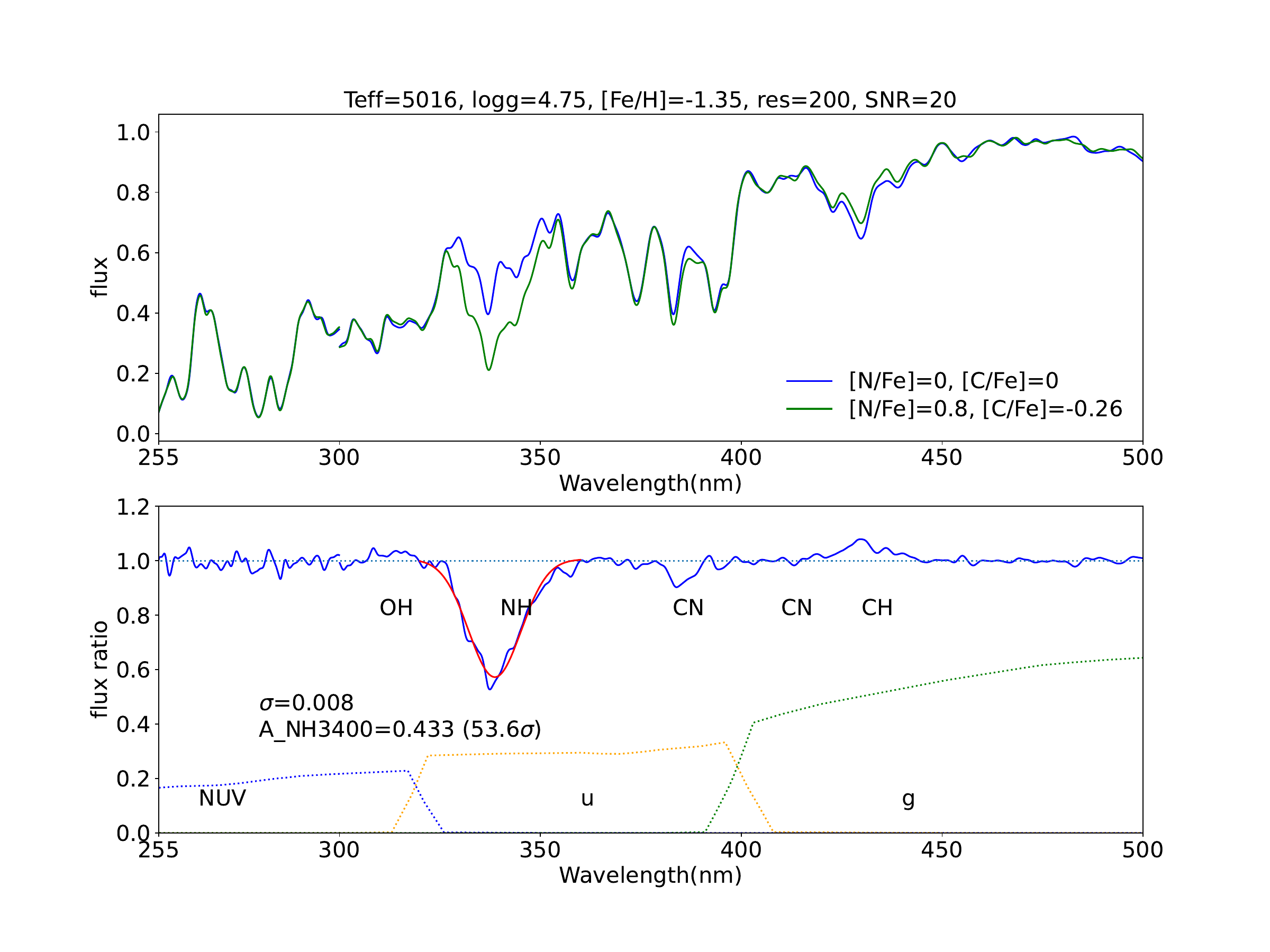}}
    \subfigure[SNR = 10]{
    \label{fig:dwarf_spectral_signal_SNR10}
    \includegraphics[width=0.8\textwidth]{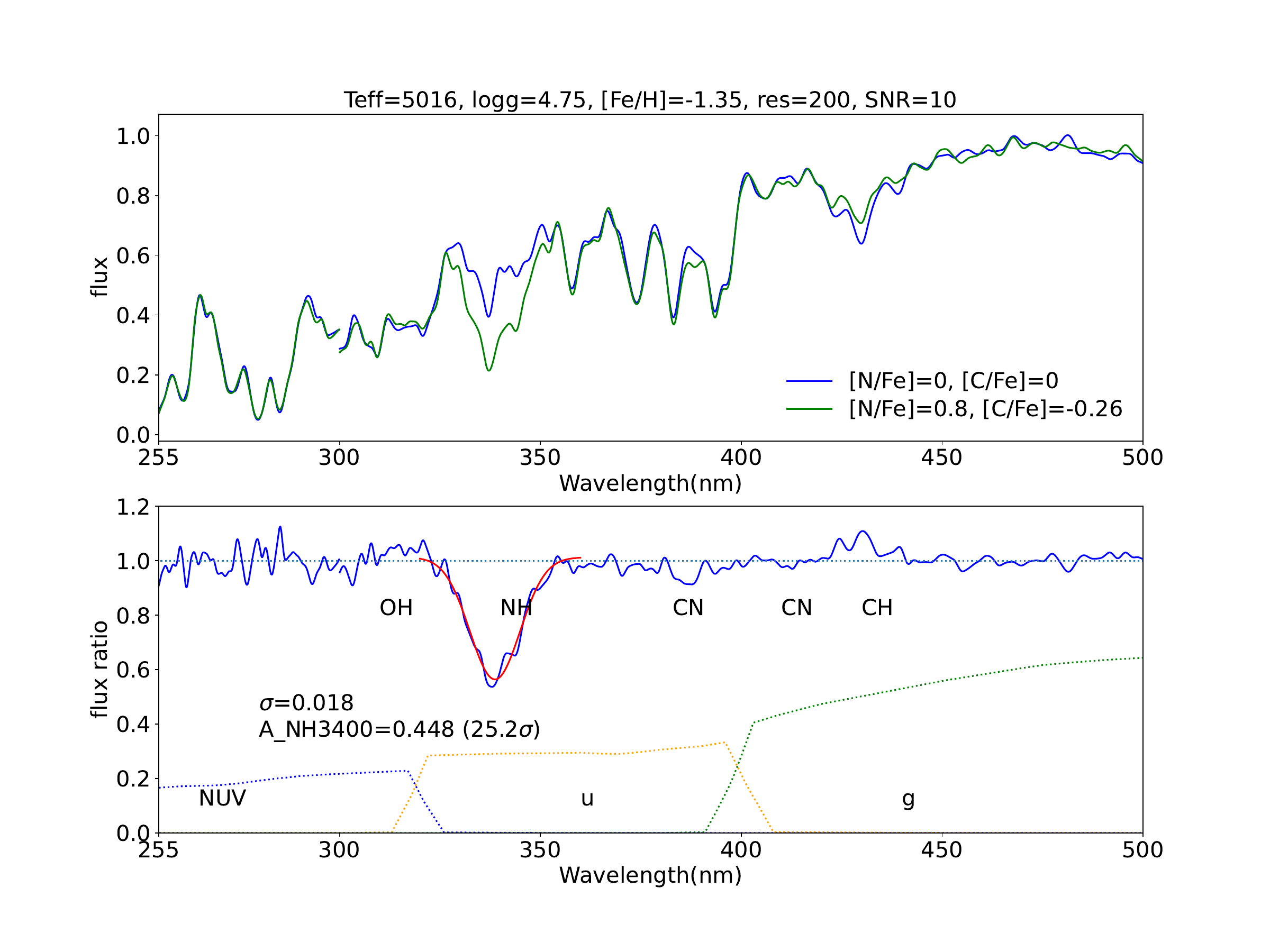}}
    \caption{Synthetic spectra with SNR of 20 (a) and 10 (b) for a representative dwarf star. See the text for more details.}
    \label{fig:dwarf_spectral_signal}
\end{figure*}

We did the same for a representative star from the dwarf star sample (Figure \ref{fig:dwarf_spectral_signal}). Compared to the giant stars, only NH3400 feature is significantly different between a N-rich star and a normal star. The significant level of NH3400 is 53.6$\sigma$ under an SNR of 20 and 25.2$\sigma$ under an SNR of 10, both of which are larger than 5$\sigma$. This suggests that a spectrum with an SNR of 10 is sufficient for separating N-rich dwarf stars from normal dwarf stars.

Given that N-rich stars can be identified in slitless spectra with a given SNR, there is concern about how to identify them from tens of millions of spectra. There are multiple methods to carry out such investigation: 1. one can calculate the spectral indices related to N feature lines, and select those with extreme values. This method is carefully outlined in \cite{Tang2019,Tang2020}; 2. one can find out the N-rich stars with strong N-related features by applying the outlier detection algorithm on slitless spectra with similar stellar parameters; and etc.

\section{Discussion}\label{sec:discussion}
\begin{figure*}
\center
\includegraphics[scale=0.7]{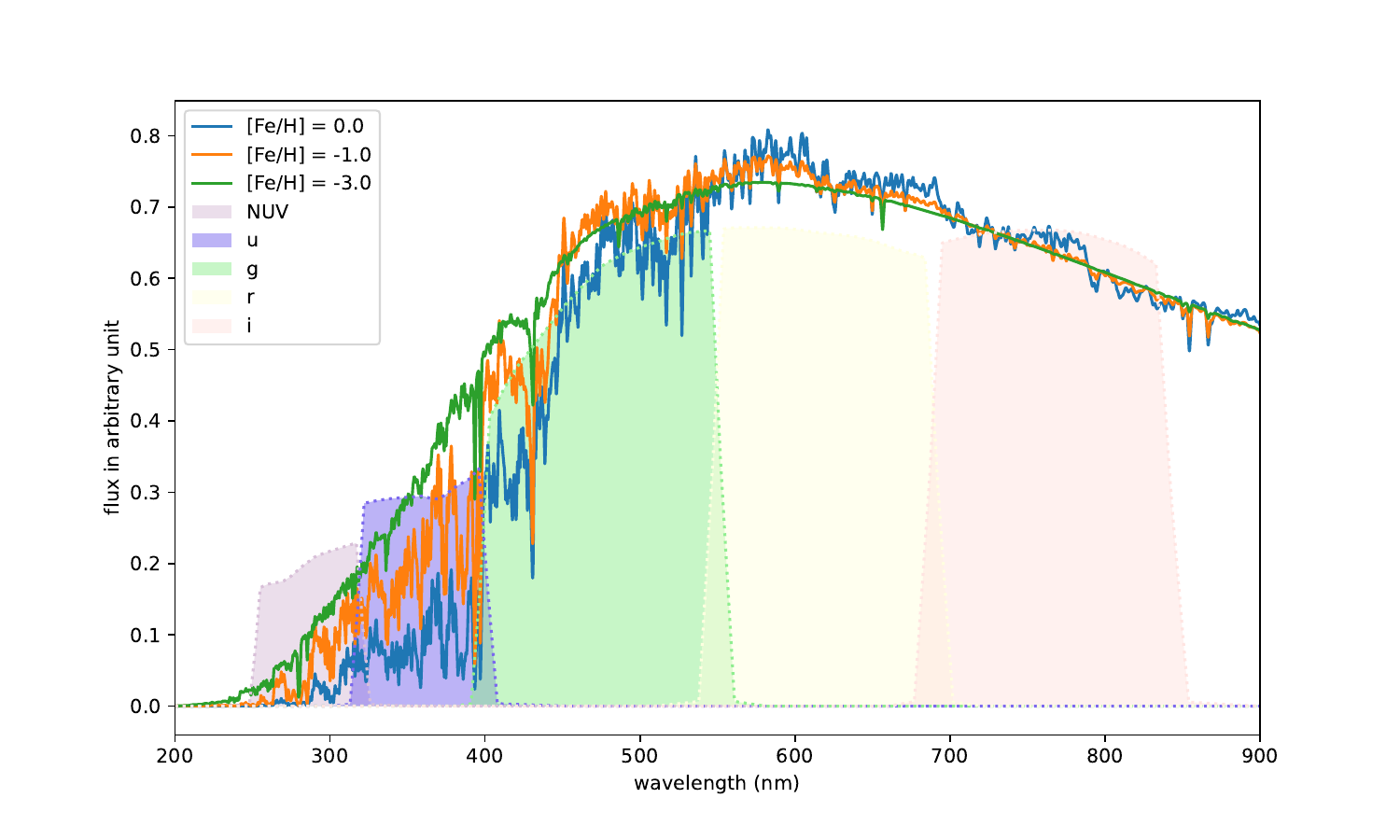}
\caption{Three synthetic spectra for three red giant stars with same effective temperature and surface gravity but different metallicity. The effective temperature is 4700 K, and the surface gravity is 1.5. The blue, orange, and green lines represent metallicity equal to 0.0, -1.0, -3.0, respectively. The spectra come from the PHOENIX synthetic library \citep{Husser2013}. Three spectra are normalized using the flux ranging from 700 nm to 750 nm. The shadow areas with different colors represent the throughput of CSST filters. \label{fig:PHOENIX_spectra}}
\end{figure*}

There are two practical issues on photometric identification of N-rich stars. The first one is how accurate can we measure the metallicity from the CSST main survey photometry? Trained with artificial intelligence, Haibo Yuan (private communication) suggested that the retrieved metallicity accuracy is $\sim$0.1 dex for FGK solar-scaled abundance stars, which is less than the metallicity bin width of 0.2 dex used in photometric identification. However, metallicity is derived based on its strong impact on \emph{u}-band magnitudes, which is also affected by N abundances. Thus, there is concern about the degeneracy between deriving metallicity and N abundances based on \emph{u}-band magnitudes. To break this degeneracy, filters that are sensitive to either metallicity or N abundances could be helpful. Theoretically, the \emph{NUV} band should be more sensitive to metallicity compared to the \emph{u} band, because of the numerous metal lines in the former band. To verify this, we retrieve spectra from the PHOENIX synthetic library\footnote{https://phoenix.astro.physik.uni-goettingen.de/} \citep{Husser2013}. Figure \ref{fig:PHOENIX_spectra} shows the flux differences in \emph{NUV} and \emph{u} band filters for three red giant stars with different metallicities. The \emph{NUV}-band magnitude is clearly sensitive to [Fe/H], but not to [N/H], given that there are no strong N-related features in this band (see Figure \ref{fig:giant_spectral_signal} and \ref{fig:dwarf_spectral_signal}). Therefore, the aforementioned degeneracy can be broken with \emph{NUV}-band magnitude or related colors. However, the relatively poor studies in oscillator strength of absorption features in this band prevent us from further investigation. Besides, it is also beyond the scope of this paper. Hopefully, the upcoming CSST photometric and spectroscopic observations in the \emph{NUV} band can provide calibration to the theoretical spectra. Since most metal-poor halo stars are alpha-enhanced, we further discuss if \emph{u}-band photometry is strongly affected by [alpha/Fe]. Following the procedure above, we retrieved two spectra from the PHOENIX synthetic library with [alpha/Fe] = 0.2 and 0.4 at $T_{\rm eff}$=4700 K, $\log g$=1.5, [Fe/H]=-1.0. We find that the magnitude difference in the \emph{u}-band is negligible ($\sim$0.005 mag). Therefore, alpha-enhancement does not affect the [N/H] determination.

\begin{figure}
\center
\includegraphics[scale=0.16]{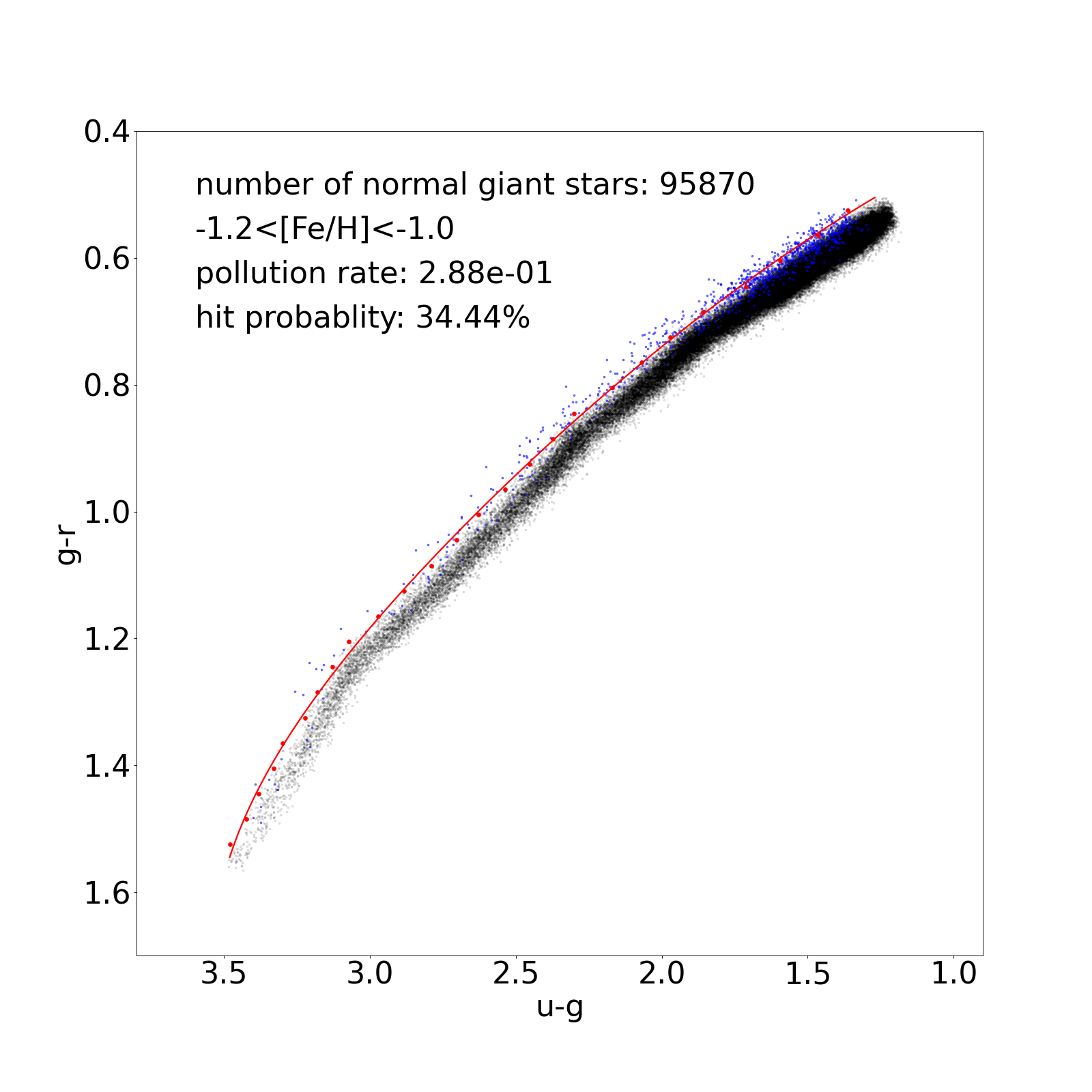}
\caption{\emph{u-g} versus \emph{g-r} for giant stars with photometric error at \emph{g} $<18.5$ mag and -1.2 $<$ [Fe/H] $<$ -1.0. Symbol meanings are the same as Figure \ref{fig:giant_ugvsgr_feh_m12m10}. \label{fig:giant_ugvsgr_feh_m12m10_phot_err}}
\end{figure}

\begin{figure}
\center
\includegraphics[scale=0.16]{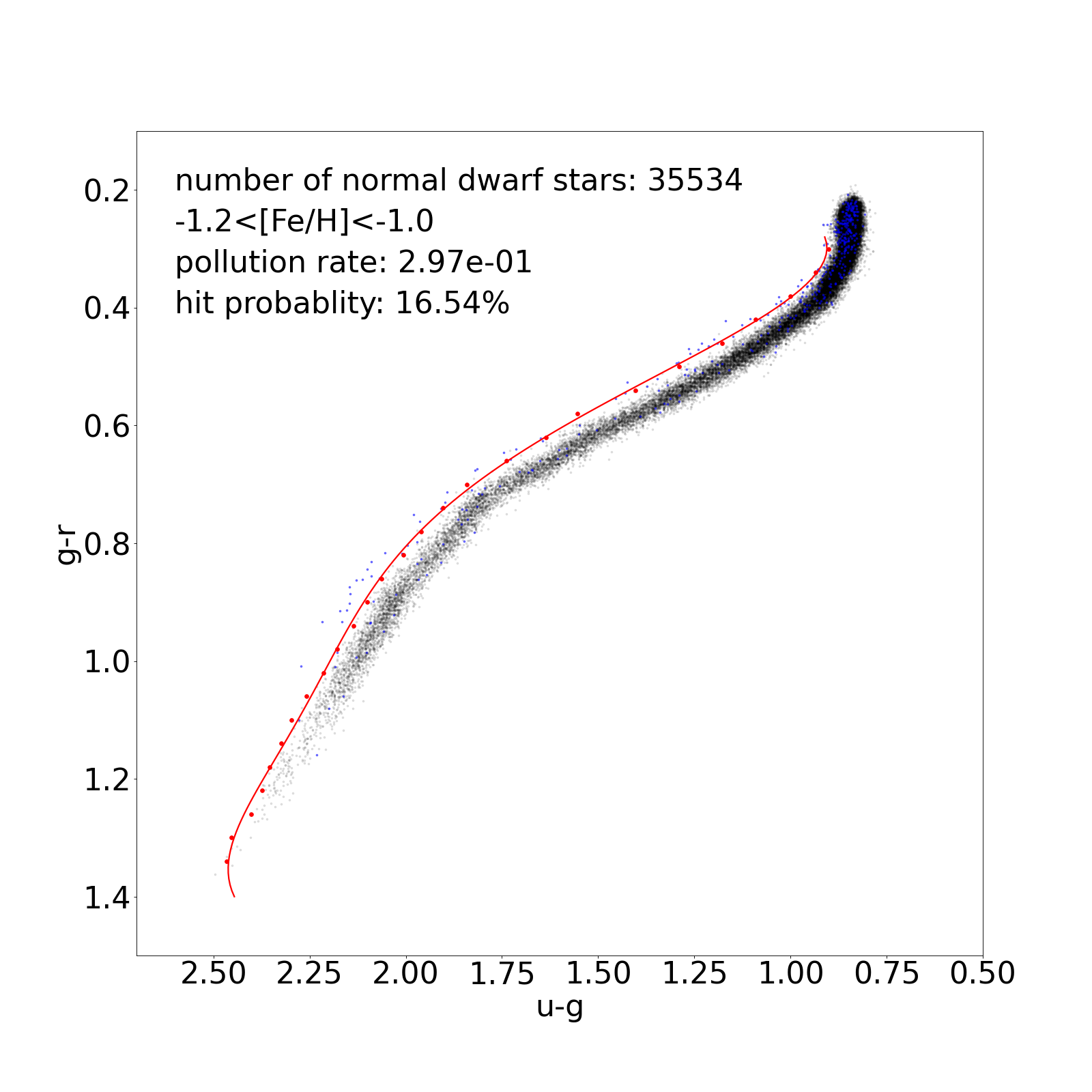}
\caption{\emph{u-g} versus \emph{g-r} for dwarf stars with photometric error at \emph{g} $<19.0$ mag and -1.2 $<$ [Fe/H] $<$ -1.0. Symbol meanings are the same as Figure \ref{fig:ugvsgr_feh_m12m10}. \label{fig:ugvsgr_feh_m12m10_phot_err}}
\end{figure}

The second one is the impact of photometric error. We added photometric errors of \emph{u, g, r} bands for all the stars in our catalog according to \cite{Qu2023}. After adding photometric errors, the efficiency of identifying N-rich stars from normal stars drops significantly due to the large photometric errors of fainter stars. To avoid large photometric errors that blur the boundaries, we limited our sample to brighter stars. In the end, we find \emph{g} = 18.5 mag achieves a good balance between maximizing sample size and minimizing  pollution rate (Table \ref{tab:table_giant_photerr} and \ref{tab:table_dwarf_photerr}).
In these two tables, we found reasonable hit probabilities and pollution rates, suggesting that the method is valid for bright stars (\emph{g} $<$ 18.5 mag).
The color-color diagrams of \emph{u-g} versus \emph{g-r} are shown in Figure \ref{fig:giant_ugvsgr_feh_m12m10_phot_err} and \ref{fig:ugvsgr_feh_m12m10_phot_err} between the metallicity bin of [-1.2, -1.0] at \emph{g} $<$ 18.5 mag. Though 18.5 mag may seem bright in photometric surveys, it is in fact fainter than any currently identified N-rich field stars, as spectroscopic observations are required.

\begin{table*}
\centering
\caption{The Pollution Rate, Hit Probability, and Coefficients of Division Lines for Giant Stars with Photometric Error.}
\label{tab:table_giant_photerr}
\begin{threeparttable}
\resizebox{\textwidth}{!}{
\renewcommand\arraystretch{2}
\begin{tabular}[b]{lllllllll}
\hline
\hline
[Fe/H] bin & pollution rate & hit probability & $a_{0}$ & $a_{1}$ & $a_{2}$ & $a_{3}$ & $a_{4}$ & $a_{5}$\\
\hline
-1.2$\rm <[Fe/H]<$-1.0 & 2.90e-01 & 34.44\% & -2.011 & 11.076 & -13.419 & 10.607 & -4.251 & 0.626\\
-1.4$\rm <[Fe/H]<$-1.2 & 2.78e-01 & 34.79\% & 5.176 & -31.625 & 84.673 & -98.796 & 55.028 & -11.881\\
-1.6$\rm <[Fe/H]<$-1.4 & 2.89e-01 & 18.01\% & 5.992 & -36.954 & 97.916 & -114.788 & 64.398 & -14.028\\
-1.8$\rm <[Fe/H]<$-1.6 & 2.75e-01 & 13.16\% & 6.545 & -39.944 & 103.637 & -119.564 & 65.977 & -14.130\\
-1.8$\rm <[Fe/H]<$-1.0 & 2.99e-01 & 17.17\% & -1.361 & 7.587 & -6.270 & 3.527 & -0.848 & -0.011\\
\hline
\end{tabular}}
\end{threeparttable}
\raggedright{\hspace{2cm}Note: $y = a_{0} + a_{1}x + a_{2}x^{2} + a_{3}x^{3} + a_{4}x^{4} + a_{5}x^{5}$, $x = $ \emph{g - r}, $y = $ \emph{u - g}.\\}
\end{table*}

\begin{table*}
\centering
\caption{The Pollution Rate, Hit Probability, and Coefficients of Division Lines for Dwarf Stars with Photometric Error.}
\label{tab:table_dwarf_photerr}
\begin{threeparttable}
\resizebox{\textwidth}{!}{
\renewcommand\arraystretch{2}
\begin{tabular}[b]{lllllllll}
\hline
\hline
[Fe/H] bin & pollution rate & hit probability & $a_{0}$ & $a_{1}$ & $a_{2}$ & $a_{3}$ & $a_{4}$ & $a_{5}$\\
\hline
-1.2$\rm <[Fe/H]<$-1.0 & 2.92e-01 & 16.91\% & 3.495 & -22.774 & 68.750 & -86.034 & 49.632 & -10.871\\
-1.4$\rm <[Fe/H]<$-1.2 & 2.14e-01 & 27.50\% & 3.745 & -24.945 & 75.017 & -94.771 & 55.338 & -12.251\\
-1.8$\rm <[Fe/H]<$-1.6 & 2.50e-01 & 30.00\% & -1.513 & 27.416 & -128.157 & 285.045 & -287.176 & 107.230\\
-1.8$\rm <[Fe/H]<$-1.0 & 3.01e-01 & 17.83\% & 3.408 & -22.078 & 66.615 & -83.273 & 48.060 & -10.544\\
\hline
\end{tabular}}
\end{threeparttable}
\raggedright{\hspace{2cm}Note: (1) We do not list relative information at the metallicity bin of [-1.6, -1.4], because the pollution rate is high when the quantile is 100\% at \emph{g} $<$ 18.0 mag. The range of application of the division line is 0.28 $<$ \emph{g-r} $<$ 1.15 at the metallicity bin of [-1.4, -1.2] and 0.28 $<$ \emph{g-r} $<$ 0.9 at the metallicity bin of [-1.8, -1.6], while for other metallicity bins, the range is 0.28 $<$ \emph{g-r} $<$ 1.4.\\ (2) $y = a_{0} + a_{1}x + a_{2}x^{2} + a_{3}x^{3} + a_{4}x^{4} + a_{5}x^{5}$, $x = $ \emph{g - r}, $y = $ \emph{u - g}.\\}
\end{table*}

\section{Conclusion} \label{sec:conclusion}

GCs are one of the oldest stellar systems in our MW, which witnessed the formation and evolution of the latter. During their co-evolution, GC stars could escape to the field, and retreating these stars is crucial for depicting the details of MW evolution. The N-rich nature of chemically enhanced populations in GCs offers a bright future for identifying GC escapees with \emph{UV} photometry and spectroscopy, e.g., CSST.
Based on the mock photometric catalogue of MW halo stars, we evaluated the efficiency of identifying N-rich stars for giants and dwarfs, separately. We used {\sc iSpec} to generate spectra with various chemical patterns, and added Poisson noise to mimic observed spectra with SNR of 10 and 20. We demonstrated that slitless spectra ($R \sim 200$) with a SNR of 10 are sufficient to identify N-related molecular features with significance greater than 5$\sigma$, and thus able to separate N-rich stars from normal stars.
In parallel, we generated spectra with different chemical patterns and convolved them with CSST main survey filters to simulate their magnitude differences. We found that the color-color diagram of \emph{u-g} versus \emph{g-r} is a potential way to identify N-rich stars, since \emph{u} band covers the strong NH and CN molecular features.

Based on our mock catalogue, we will identify a much larger number of N-rich stars than before. This large sample with a unprecedented size will be a comprehensive sample to test different MW formation scenarios, e.g., results from N-body simulations. For example, \cite{Gieles2023} evaluated the mass loss rates of GCs with stellar-mass black holes, where they found an agreement for density profiles of N-rich (field) stars between models and observations (see their Figure 12), confirming that these stars originated from GCs. In the near future, the largest sample of N-rich (field) stars identified using CSST photometry will reveal their spatial distribution, velocity distribution and even chemical abundances in a statistically significant level. Further combination with state-of-the-art N-body simulations will be a promising to address many details during the MW evolution.

\begin{acknowledgements}
We thank Haibo Yuan for helpful discussions. We thank the anonymous referee for insightful comments. J.Z. and B.T. gratefully acknowledge support from the China Manned Space Project No. CMS-CSST-2021-B03, the National Natural Science Foundation of China under grant No. 12233013, 12073090, and the Natural Science Foundation of Guangdong Province under grant No. 2022A1515010732. J.G.F-T gratefully acknowledges the grant support provided by Proyecto Fondecyt Iniciaci\'on No. 11220340, and also from ANID Concurso de Fomento a la Vinculaci\'on Internacional para Instituciones de Investigaci\'on Regionales (Modalidad corta duraci\'on) Proyecto No. FOVI210020, and from the Joint Committee ESO-Government of Chile 2021 (ORP 023/2021), and from Becas Santander Movilidad Internacional Profesores 2022, Banco Santander Chile.
\end{acknowledgements}
\bibliography{ms2023-0231Ref}
\bibliographystyle{aasjournal}

\appendix
\section{Additional figures}

\begin{figure*}
    \centering
    \includegraphics[width=0.45\textwidth]{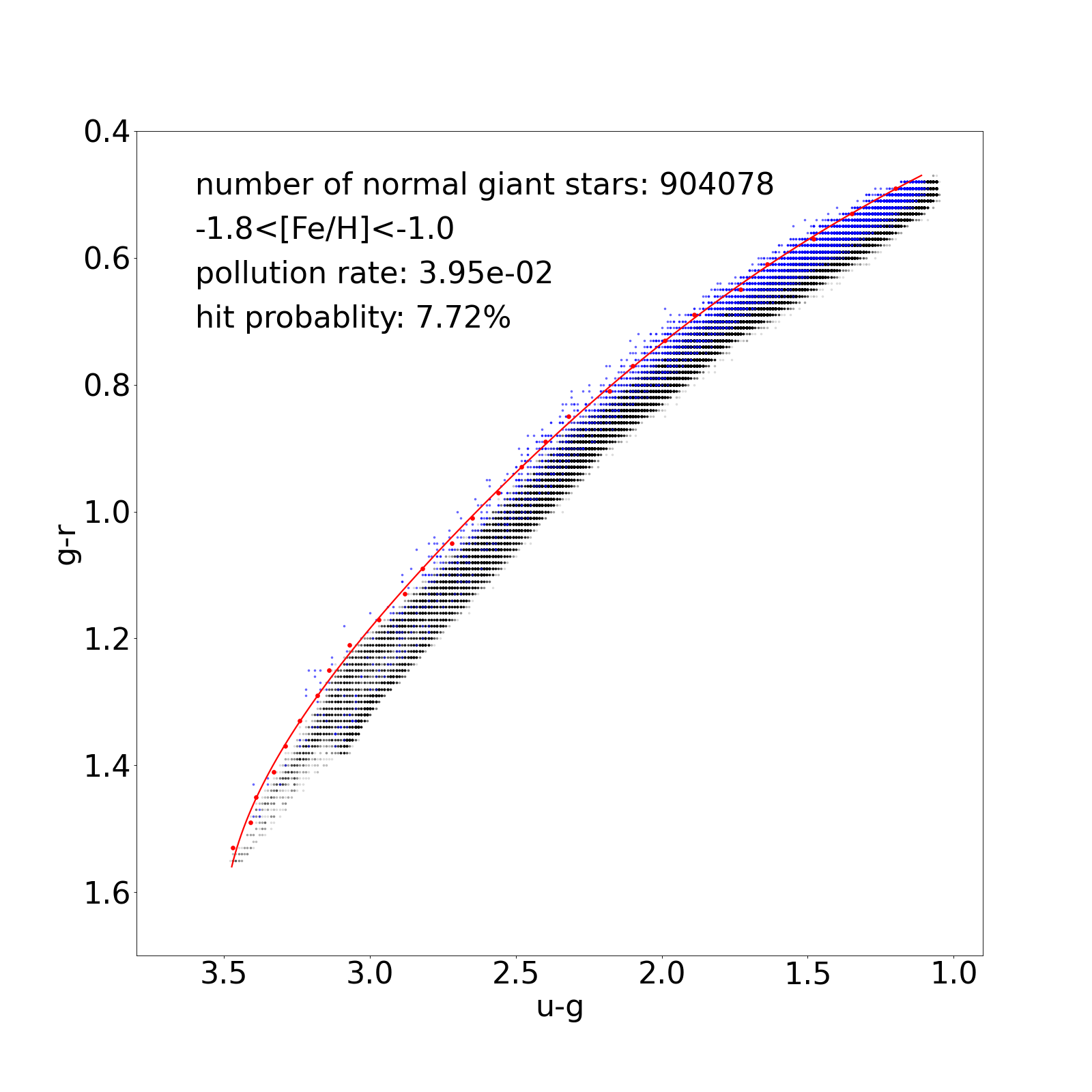}
    \includegraphics[width=0.45\textwidth]{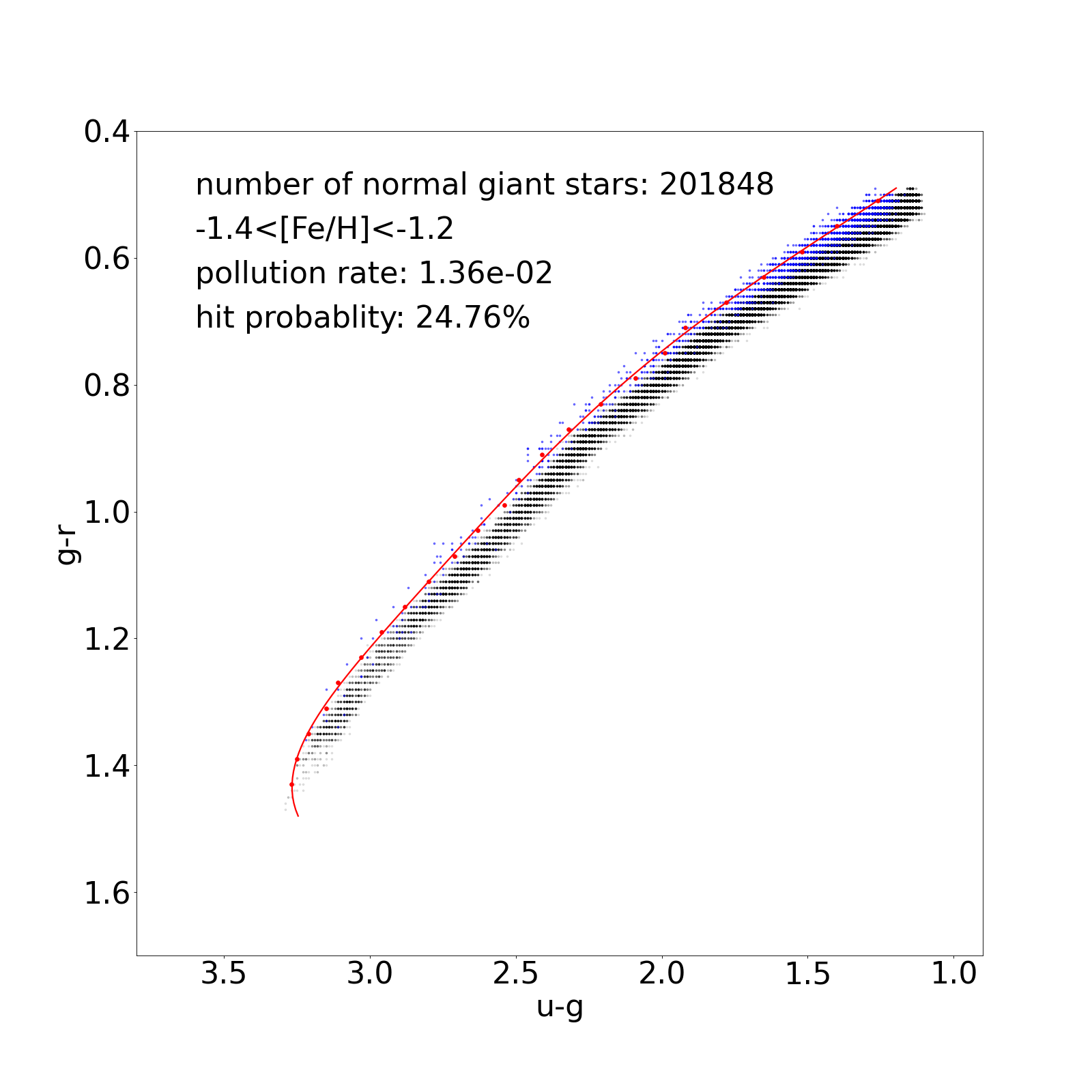}
    \includegraphics[width=0.45\textwidth]{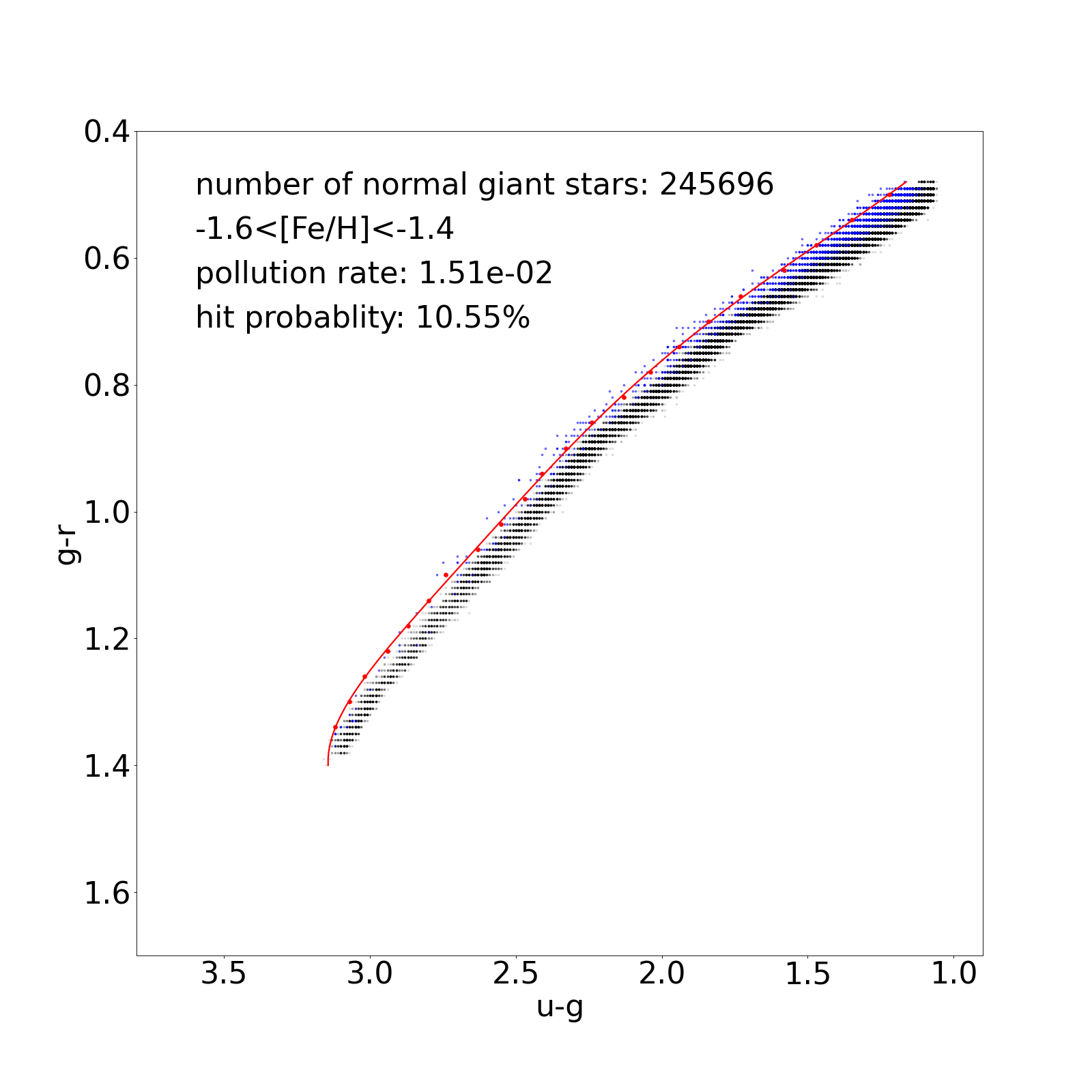}
    \includegraphics[width=0.45\textwidth]{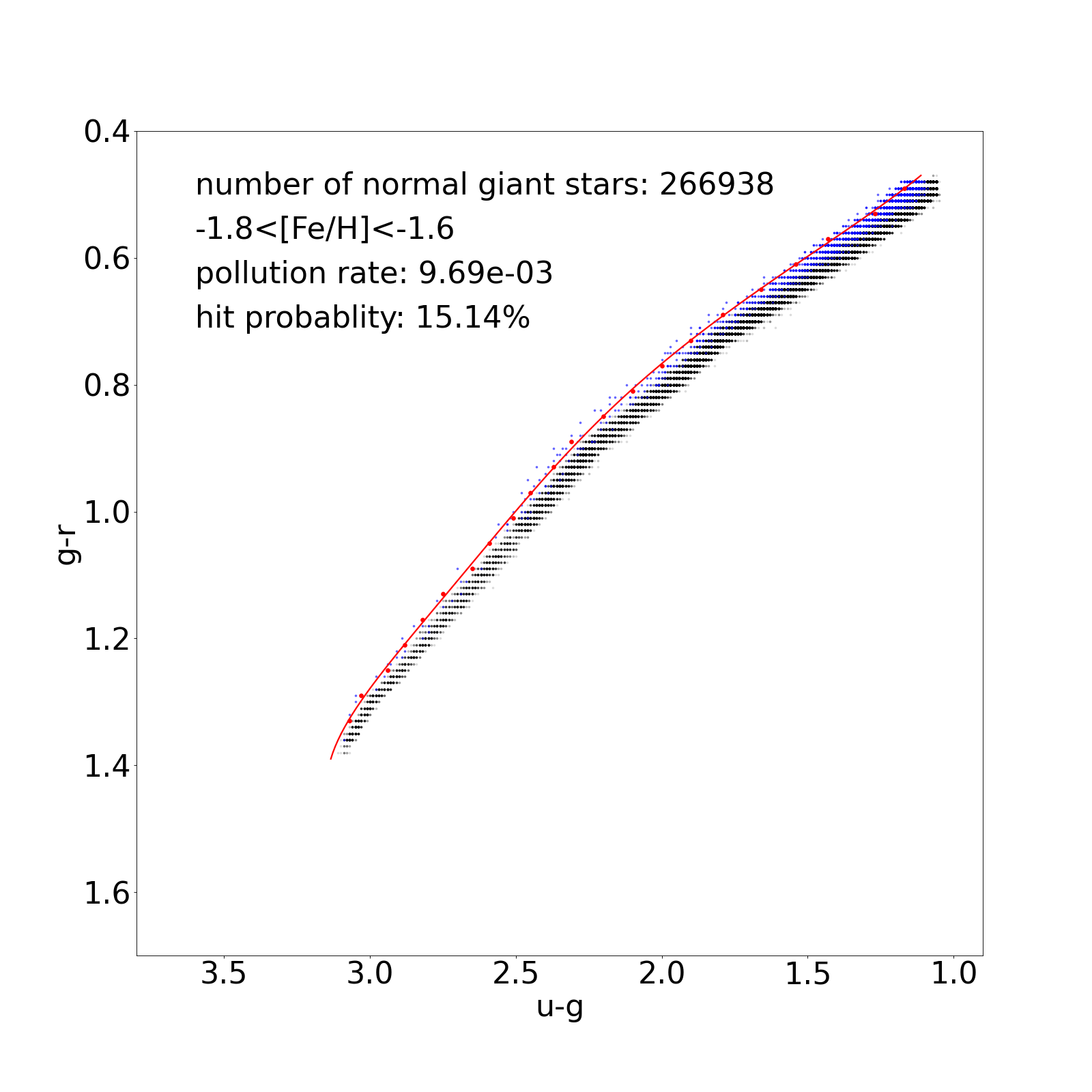}
    \caption{\emph{u-g} versus \emph{g-r} for N-rich giant stars (blue) and normal giant stars (black) at different metallicity ranges. Symbol meanings are the same as Figure \ref{fig:giant_ugvsgr_feh_m12m10}.}
    \label{fig:nhgiant_ugvsgr_otherfeh}
\end{figure*}

\begin{figure*}
    \centering
    \includegraphics[width=0.45\textwidth]{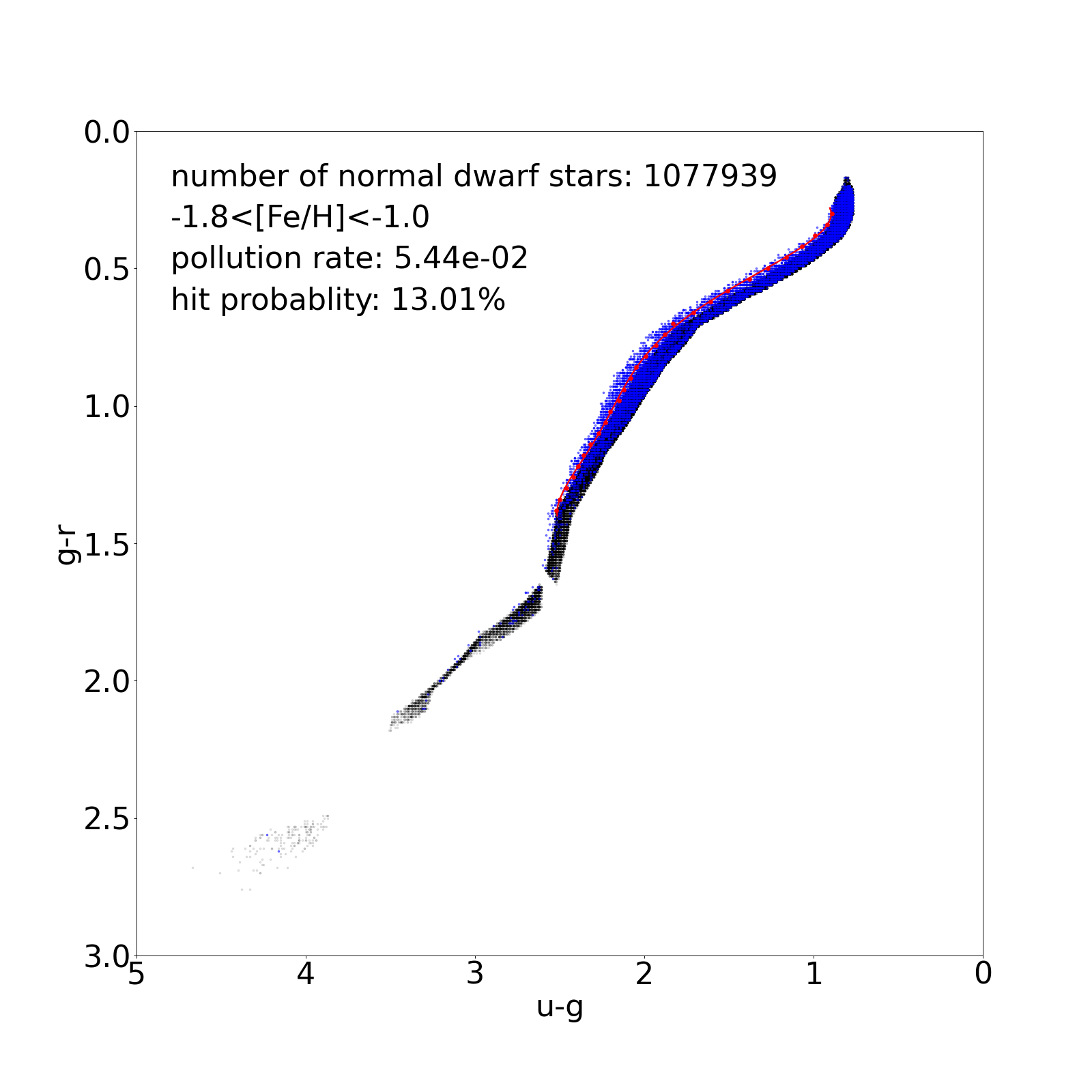}
    \includegraphics[width=0.45\textwidth]{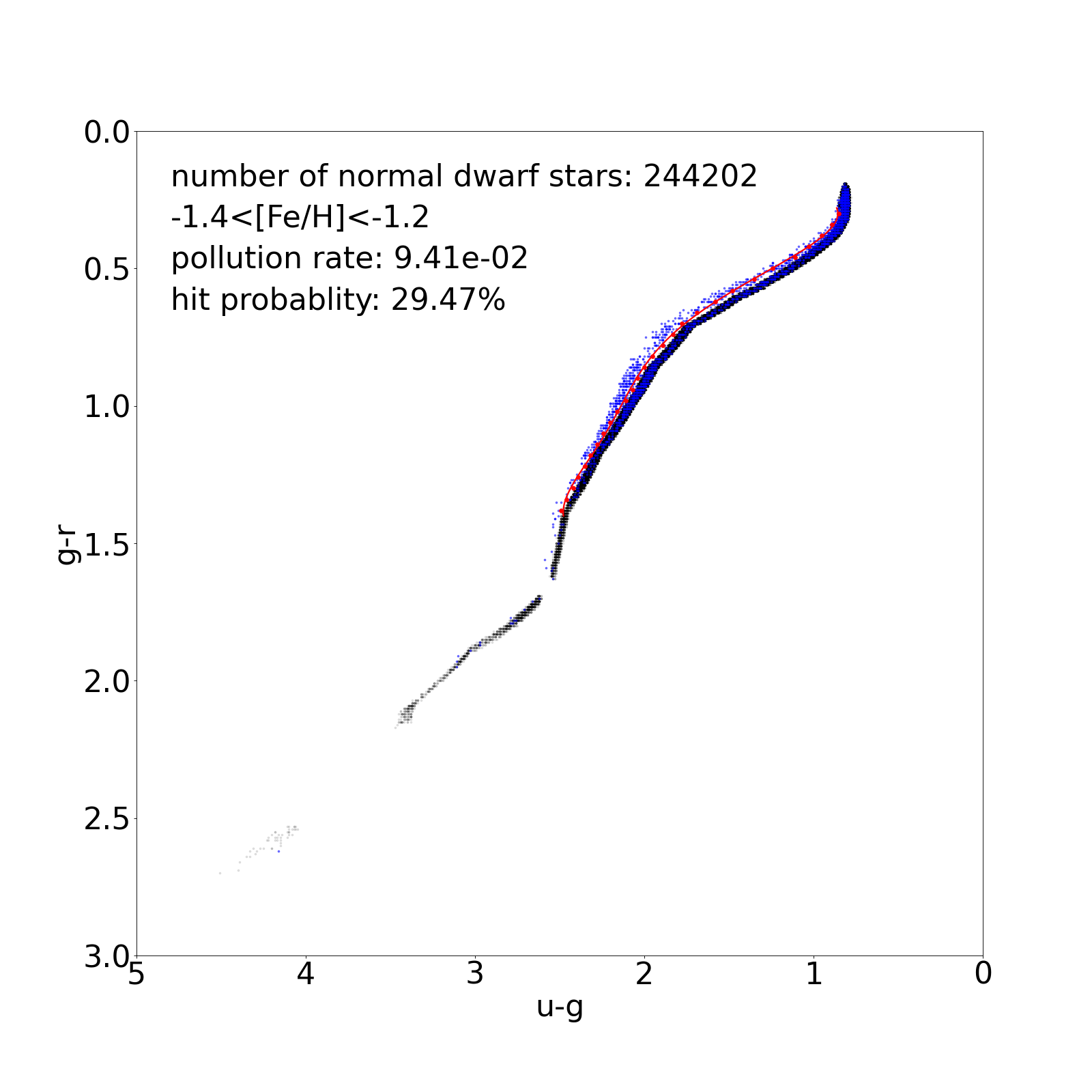}
    \includegraphics[width=0.45\textwidth]{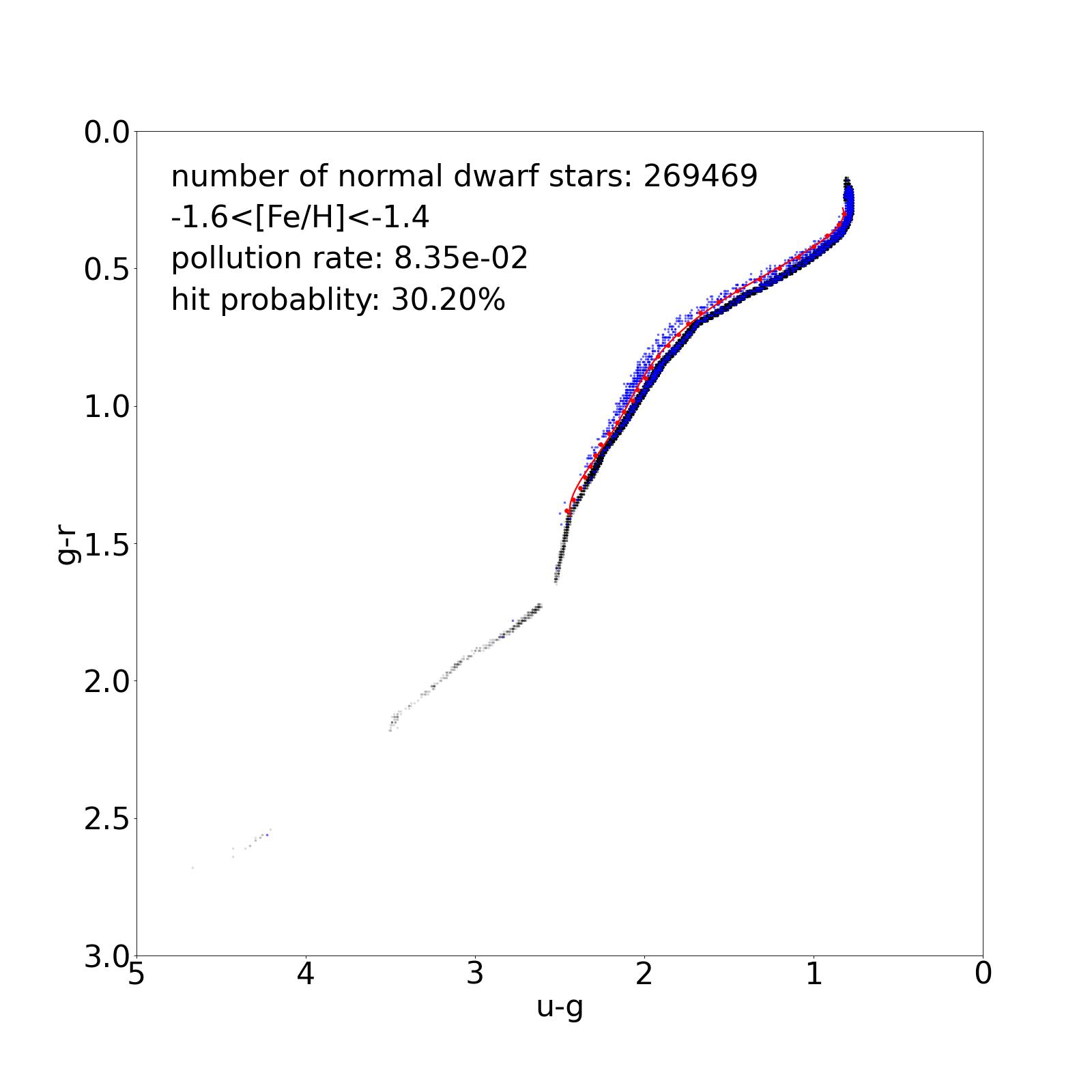}
    \includegraphics[width=0.45\textwidth]{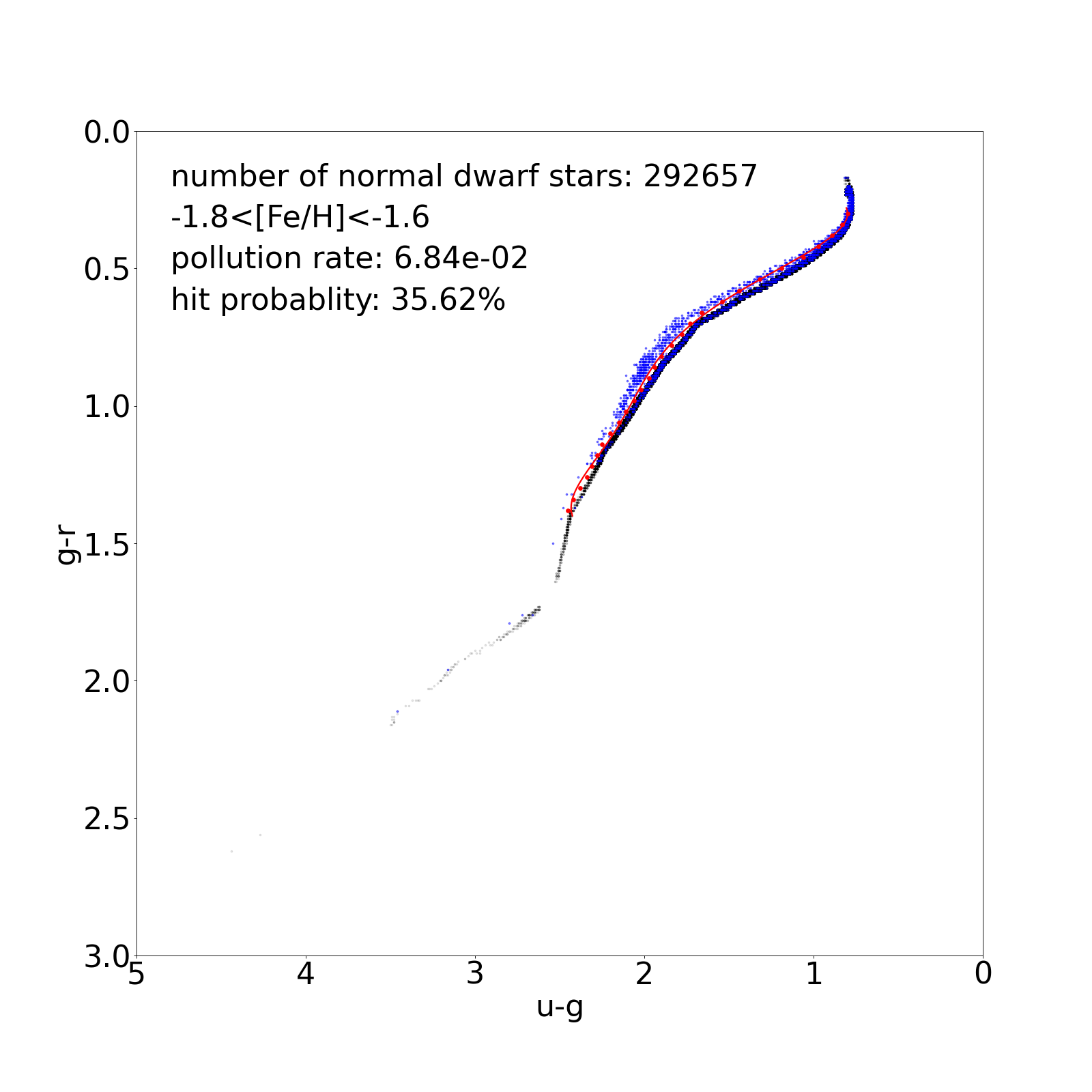}
    \caption{\emph{u-g} versus \emph{g-r} for N-rich dwarf stars (blue) and normal giant stars (black) at different metallicity ranges. Symbol meanings are the same as Figure \ref{fig:ugvsgr_feh_m12m10}.}
    \label{fig:ugvsgr_otherfeh}
\end{figure*}
\end{document}